# 修士学位論文

Analysis of Anomalies in the Internet Traffic Observed at the

Campus Network Gateway

キャンパスネットワークゲートウェイにおける

異常インターネットトラフィックの分析

平成 ２２年度

総合分析情報学コース

096404

Veronica del Carmen Estrada

ヴェロニカ　デル　カルメン　エストラダ

指導教員 中尾彰宏 准教授

Analysis of Anomalies in the Internet Traffic Observed at the Campus
Network Gateway

Veronica del Carmen Estrada

Submitted to

the Graduate School of Interdisciplinary Information Studies

The University of Tokyo

on January 14, 2011

in Partial Fulfillment of the Requirements for the

Degree of Master of Interdisciplinary Information Science

Accepted by

.........................................................................................................................

Akihiro Nakao, Associate Professor

Interfaculty Initiative in Information Studies

The University of Tokyo

Thesis Supervisor

# 論 文 要 旨

Analysis of Anomalies in the Internet Traffic Observed at the Campus
Network Gateway

キャンパスネットワークゲートウェイにおける
異常インターネットトラフィックの分析

東京大学大学院

学際情報学府 学際情報学専攻

総合分析情報学コース

49-096404

Veronica del Carmen Estrada

ヴェロニカ　デル　カルメン　エストラダ

# ABSTRACT

One of the key factors to bring success to machine learning (ML) based systems in the operational environments is trying to understand the domain-specific data better. A considerable portion of the literature in ML applied to intrusion detection uses outdated data sets based on a simulated network with a limited environment. The characteristics of the real network traffic captured in a large environment include inherent diversity in the protocol behavior and the "crud" [1]. Moreover, the difficulties in capture process should not be underestimated. Flaws usually appear in datasets and the way we handle them may impact on measurements. Finally, the detection capacity of intrusion detection is highly influenced by the system configuration.

We focus on a topic rarely investigated: the characterization of anomalies in a large network environment. Intrusion Detection System (IDS) are used to detect exploits or other attacks that raise alarms. Some IDSs are able to detect another category of events that indicate that something is wrong but it may not be an attack. These anomalous events usually receive less attention than attack alarms, causing them to be frequently overlooked by security administrators. However, the observation of this activity contributes to understand the traffic network characteristics. On one hand, abnormal behaviors may be legitimate, e.g., misinterpreted protocols or malfunctioning network equipment, but on the other hand an attacker may intentionally craft packets to introduce anomalies to evade monitoring systems.

With the trend in stealthy attacks and the prevalent threat of botnets, anomalous events observed in traffic should be considered more seriously than ever. Then we need a framework of reference to understand anomalous behavior, questions that arise are how frequent these events are and what is causing them.

Recently, researchers have been enthusiastic about applying methods to automatic or semi-automatic learning in security logs. However, there is low acceptance of these techniques in the operational real-world network environment [2]. Considering that we want to achieve more efficient use of ML in the future, at this stage, we give a priority to gaining knowledge from the traffic behavior observed at the main campus gateway. This thesis analyzes problems that arise in data pre-processing and are not anywhere documented. We assess the quality of the captured data, achieve a better comprehension of the major traffic characteristics and find possible explanations for anomalies observed in the captured network traffic.

Anomalies found in operational network environments may indicate cases of evasion attacks, application bugs, and a wide variety of factors that highly influence intrusion detection performance. Then, we conduct a study to explore the nature of anomalies found in U-Tokyo Network using cooperatively Bro and Snort IDS among other resources. We analyze 6.5 TB of compressed binary tcpdump data representing 12 hours of network traffic.

Our major contributions can be summarized in: 1) reporting the anomalies observed in real, up-to-date traffic from a large academic network environment, and documenting problems in research that may lead to wrong results due to misinterpretations of data or misconfigurations in software; 2) assessing the quality of data by analyzing the potential and the real problems in the capture process.

Thesis Supervisor: Akihiro Nakao
Title: Associate Professor
Interfaculty Initiative in Information Studies,
The University of Tokyo

# Acknowledgements

To begin with, I would like to express my deeply gratitude to Associate Professor Akihiro Nakao, who gave me the opportunity of conducting this master thesis in his laboratory. I have learned innumerable things under his supervision. Also, my thanks to other professors and academic staff who make this master possible.

My stay in Japan has been an incredible life experience. I would like to give a special acknowledge to the Ministry of Education, Culture, Sports, Science and Technology of Japan for scholarship support during the master school and the Japanese language school. My thanks to the Japan Student Service Organization for all the support provided to students.

I would like to thank the University of Tokyo for the invaluable research framework and the amazing structure to support foreign students. In particular, I would like to thank the academy staff of the International Center for helping me to deal with daily life issues. Also, my thanks to Ms. Sato and Ms. Yamada who have been always kind in providing assistance with all the official papers.

I would like to thank Dr. Carlos Segura and Dr. Pablo Fierens for their tuition while I was doing research in Argentina.

My gratitude to Dr. Carlos Rosito and Professor Jorge Alberto from University of Buenos Aires and Dr. Roberto Perazzo from ITBA University for their strong support and letters of recommendations to achieve the scholarship. I also give a special thank to Carlos Benitez and my colleagues and friends at Si6 Information Security Laboratory.

I would like to express my gratitude to the developers of Bro IDS, especially to Dr. Vern Paxson and Dr. Robin Sommer for helping me to understand the inner characteristics of Bro IDS.

My gratitude is also to the open source software community for collaborative development support, and special thanks to Snort IDS developers.

I would like to thank Scott Sanner and the Australian National University for organizing the 2010 Canberra Machine Learning Summer School.

I would like to thank Iñigo Perona for the discussions about KDD features and for providing me with a script to calculate KDD features values.

My thanks to my colleagues in our group at Nakao Lab, who inspire me to work hard. Special thanks to Iwase for providing help at the first stage of the trip, to Yan for his interesting conversations and to Ando for bringing me the attention to a potential problem with data used for this research. My thanks extend to Ito and Putrama, especially for helping me in formal procedures during master school.

My thanks to the Argentine Embassy and CEGAJA Argentine Embassy Students Group for attend all the matters related as my life as a foreigner. My warmly thanks to Former Argentina Ambassador, Daniel Polsky, for helping me in a delicate situation.

I warmly thank Ms. Iiri from the Japanese Volunteer Program. She opened her home to introduce me the Japanese culture and patiently answered me all my doubts about Japanese grammar.

My thanks to my language teachers Silvana and Daniel for the interesting classes to improve my communication ability before coming to Japan.

Life in another country would have been very hard without TIEC, Mr. Tanaka and Mr. Sakamoto and the Chiyoda Group. They always organized activities to help understanding the Japanese society. I met many of my friends in these activities.

Finally, my deepest thanks to my parents for bringing me into this world, to my brother for being protective, to my sisters for providing me with plenty of resources during my childhood; my thanks to Rudi for his love and tolerance; my sincere thanks to Graciela and Leandro for their extreme generosity and for the strong encouragement to do this trip; my thanks to Eduardo and my nephews for their special travel gift; and thanks to Feni, Caro, Flor, There, Tracy, Kayo, Louise and Kragen for rescuing me from my despair when things were darkest.

# Table of Contents

# List of Figures

# List of Tables

# Chapter 1

# Introduction

## 1.1 Background

Computer networks are exposed to different types of threats that could affect the integrity of the data transmitted, confidentiality of the information and/or the availability of network services. Network analysis and intrusion detection are countermeasures to thwart botnets, viruses, spam, worms and any other types of malicious attacks on the network. To illustrate, the number of intrusion attempts per day in 2003 was in the order of 25 billions [3]. Google's researchers say that even a single visit to an infected web site enables the attacker to detect vulnerabilities of the user's applications and force the download a multitude of malware binaries [4]. Metrics on web-based malware presented in the cited work reveal with high confidence that 10 percent of the 4.5 million analyzed URL are malicious. From an operational point of view, the increasing trend in the amount of security controls deployed in networks is indicating that everyday intruders baffle researchers with new insidious attacks. Even though old threats persist, techniques evolved: greed for notoriety, naïve viruses and overt attacks became less attractive. Economy factors and activism are motivating the interest in stealth and complex attacks.

Botnets have recently become popular as the platform from which to launch other attacks, such as Distributed Denial of Service (DDoS) or spam. With covertly installed malware, an attacker (*botmaster*) can remotely control an infected machine (*bot or zombie*) using different communication channels, in a centralized (IRC and HTTP) or distributed architecture (P2P). According to Symantec a significant spike in new threats activity occurred during 2008. This company observed an average of 75,158 active bot-infected computers per day in 2008, an increase of 31% from the previous period. A 90% of the spam activity is generated in bot networks. Their population is dynamic over time making difficult to measure their size and estimate their prevalence. Bots can be inactive for several months or hours. Botmasters use techniques such as bot cloning or bot migration to avoid being tracked and make measurements difficult [5]. Vint Cerf, who is considered by the community as one of the "fathers" of Internet, told to the audience of the World Economic Forum 2007 that approximately 600 million computers are connected to the Internet, and that 150 million of them might be participants in a botnet.

Research on pattern recognition and computational intelligence may be to the rescue against such menaces in the Internet. However, research on unconventional attacks recognition is scarce and so is integration of different methods.

Anomalies are part of the Internet [6]. Anomalies detection is conducted using different approaches in response to the different goals that researchers try to achieve. Network security study has been developing methods for recognizing malignant patterns. On the other hand, network traffic study has been investigating methods for traffic classification mainly for traffic engineering. In a previous published paper [7], we observe that traffic classification techniques developed for traffic engineering could be combined into the current study of intrusion detection even further. A caveat is the volume of traffic logs that can be overwhelming in some high-speed networks. Some methods to filter out uninteresting data could alleviate the job of intrusion detectors.

## Machine Learning Applied to solve Information Security Problems

Due to the necessity of analyzing huge amounts of data produced by one or more sensors deployed in large network environments, machine learning techniques rise in popularity among intrusion detection researchers. Such methods give the possibility of devising a hybrid system aiming to identify useful information for the two groups aforementioned. We observed two main trends: clustering sensor's alerts and detecting outliers in "normal" behavior profiles.

A major goal in machine learning research is the automatically or semi-automatically recognition of complex patterns in data. Algorithms are trained with data samples to learn characteristics in order to make intelligent decisions using new data. In a nutshell, a machine learning project includes all or some of this phases: data preparation or pre-processing, dimensionality reduction, training, evaluation and visualization. More information on these crucial steps is presented below.

- *The quality of the sample is critical to achieve an unbiased representation of the domain and should be adequately pre-processed prior the training process.*

The landmark study on data preparation is the work of Lee and Stolfo [8], who established a framework for Mining Audit Data for Automated Models for Intrusion Detection (MADAM ID). Based on this framework, data mining algorithms were applied to the 1998 DARPA Intrusion Detection Evaluation showing good results.

DARPA data set [9, 10], which was collected and distributed by the MIT Lincoln Department, was the first standard corpora for evaluation of computer network intrusion detection systems. Data was collected in a simulated military network environment and includes a wide variety of intrusion attacks.

Stolfo prepared KDD Cup 1999 data set [11], which was built based on DARPA'98. Training set contains 4,900,000 connection vectors, each of which contains 41 features and a label indicating if it represents normal traffic or an attack. Simulated attacks fall in four different categories: Denial of Service Attack (DoS), Use to Root Attack (U2R), Remote to Local Attack (R2L) and Probing Attack. KDD features are also classified in four groups: basic features (9) that are derived from packet headers, content features (13) that are derived payload of TCP packets, time based features (9) to inspect the packets within a temporal window of 2 seconds and host based features (10) to inspect the packets within a window of one hundred connections.

Some alternatives to the DARPA and KDD benchmarks are the Genome Data, attacks obtained through honeypots and sandbox, data simulation, attack injection in normal traffic, and dump real network data placing confidence in the knowledge of network and security experts to classify attacks. Each of the presented solutions has limitations.

- *The selection of the algorithm should be carefully considered to achieve project success.*

Machine learning is a scientific interdisciplinary field. Proposed ideas may come from several areas such as probability theory, statistics, pattern recognition, cognitive science, data mining, adaptive control, computational neuroscience and theoretical computer science. Three main approaches are supervised learning, unsupervised learning and reinforcement learning.

Finding the appropriate algorithm for the domain of study is not simple and researchers have to consider the trade-off between two or more variables, such as algorithm performance in terms of correct classifications, computer resources consumption, ability to treat nominal features and ability to learn in unbalanced samples among other algorithms specifics.

Self-organizing maps in addition to other types of neural networks, support vector machines, decision tree and K-means are some examples of algorithms that require attribute vectors for training and testing. Each input instance is characterized by the value of attributes or features that measure different aspects of the instance [12]. The use of public available datasets that contains classified and labeled instances of attacks can reduce time.

- *Weak validations may undermine the efforts made along long-term research projects.*

The methodology used for validation depends on the approach of machine learning research. Classification methods may be served by pre-labeled data while clustering methods may use cross-validation. The simplest way to evaluate the performance of the algorithm is by counting the proportion of correctly predicted samples. For this purpose, one approach is to subdivide the dataset in training and test samples that are mutually independent [13]. Results obtained in the evaluation using previously unseen samples may be use as a feedback to improve the training phase.

## 1.2 Motivation and Challenges

We trust on the ability of machine learning algorithms to leverage the power detection of standard IDS deployed in operational network environments. However, to solve real and interesting problems, such ability is highly influenced by external factors. First, if the quality of data were low, input samples would fail in representing the domain. Thus, the algorithm would not be able to recognize interesting patterns of data, or the presence of recognized patterns might not indicate that the system was a contribution to solving real world problems. Second, decisions made by the system designer must not be neglected. Researchers must decide the representation of raw data, select the most appropriate algorithm for data characterization, specify how this algorithm is parameterized, validate results and finally give a good interpretation of the output. Some approaches try to minimize the human-machine interaction but systems still rely on human intuition.

Moreover, data preparation usually consumes the bulk of effort invested in a learning process [12]. As a rule of thumb, it is recognized that up to 80 percent of the time spent on a machine learning project is gone during the pre-processing phase.

People involved in the DARPA and KDD evaluation had substantially contributed to the field of intrusion detection. However, flaws in the mentioned data sets caused by simulation artifacts gave reasons to criticize the project from the beginning. In 2000, the DARPA dataset received a comprehensive criticism of the failure to demonstrate the accuracy of the model [14]. Other research confirms the afore-mentioned criticism by conducting some experiments they showed that simulation errors could lead to inaccuracies in network traffic affecting anomaly detection systems that model normal behavior [15]. Moreover, the KDD dataset is not adequate to train machine-learning algorithms for misuse detection. The limitations of the dataset, detailed in [16], cause a misuse detection system to fail in the identification of 50% of the KDD's attack categories.

While researchers acknowledge the KDD problems, the use of this benchmark dataset is the solution at hand to evade some of the challenging problems mentioned at the beginning of this section. Thus, it follows that in spite of the caveats, it is not rare to find machine learning projects that still use KDD for training and/or validation. We consider that neglecting KDD issues poses a big risk to the machine learning literature applied to the information security field since the study of some topics may be biased. For instance, considering the case of features selection, a question that arises is what features should we use for reliable detection of attacks. In trying to find answers, we observe that several studies [17, 18, 19, 20, 21, 22] are based on KDD. Consequently, possible answers are limited to the list of 41 features, which in some cases is reduced to only 6 basic features. Each feature can be good as discriminating one class and bad for discriminating others. Kayacik and Heywood [19] demonstrated that three of the attack classes that make up 98% of the training KDD set are highly related to certain features that make their classification easier.

KDD is not the only problem. Indeed, there are several pitfalls affecting the way machine learning based research is conduct that result in a low acceptance of these techniques in an operative real-world network environment [2]. Zanero reviews issues on the evaluation of IDS systems and also mentions the difficulties in simulating background traffic [23]. Finally, Gates and Taylor [24] discuss challenges in the anomaly detection paradigm and closely examine assumptions commonly made in network anomaly detection.

**An assessment on true data**

Large academic networks present a unique environment that involves a wide variety of applications, protocols, and network equipment as well as heterogeneous behavior of users, administrators and intruders. For this thesis research, a 6.5 TB tcpdump binary data captured at the main link of our campus were available to us. Link operates at maximum rate of 10Gbps. Packets contain full payload and traces include uplink and downlink. A caveat is that a priori information on network layout as well as protocols found on the network is extremely scarce.

An initial approach was to extract new features that characterized U-Tokyo network traffic. To have a reference, we calculated the values for the 41 features [11] using our capture and the help of Bro IDS and a script developed for the project described in [25]. Our preliminary results showed that several values were zero. We also encounter limitations with Bro when calculating the traffic transfer in connections, particularly to measure the bytes transferred in anomalous traffic. The attempts to cluster Bro and Snort outputs using Self-organizing maps and K-means algorithms also failed. It was hard for us to apply machine learning methods efficiently. Without having references for our data, clustering methods did not help us at this preliminary stage to achieve a better comprehension of the major traffic characteristics and anomalies observed in the captured network traffic.

We changed our approach motivated by the necessity to grasp the nature of anomalies presented in Internet and to understand the roots of problems that arise in pre-processing and are not anywhere documented. We observe a wrong methodology when data that do not represent the domain is used in machine learning research projects. Key factors to bring success to machine learning based systems in operational environments are the better comprehension of the domain-specific data and avoid the treatment of security alerts with a "black box" methodology as often seen [2]. Our thesis is that the most important part of pre-processing is auditing (1) real, (2) up-to-date and (3) large amount of data before selecting reliable features for IDS.

Therefore, the outcome of this thesis serves as a baseline for future research based on machine learning methods or other techniques. Our main contributions are: 1) a summary of anomalies observed in a large academic network environment that includes behavior patterns that are not considered in KDD dataset and 2) sharing our experiences dealing with an enormous amount of packets. Given the interdisciplinary character of this area, our efforts are made toward a better understanding of anomaly detection and a greater interaction between different groups.

## 1.3   Thesis Outline

Below, there is a brief summary of the following chapters.

- Chapter 2

  Introduces the tools and methodology to evaluate the data.

- Chapter 3

  First, discusses problems related with the data captured process. Later, presents characteristics of network traffic using two approaches: connections status and amount of traffic exchange. The analysis of bogus connections ends this chapter.

- Chapter 4

  Presents the most representative cases of anomalies observed in the network. The scope of the analysis covers pathologies found at different levels of the network stack illustrating a wide variety of cases ranging from fragmentation, bad packet checksums, TCP anomalies and application anomalies.

- Chapter 5

  Concludes the thesis by summarizing our contributions and presenting interesting ideas for future work.

# Chapter 2

# Methodology

## 2.1   Stages of the Research Project

With the purpose to understand the root causes of network anomalies, we conducted qualitative and quantitative observations on the network traffic characteristics. Moreover, to observe the problematic from different angles, the background necessary for conducting experiments covers several disciplines. Thus, we accomplished this research project in four stages.

- ▪ Literature Study

  The main goal at this phase was to comprehend the context of the multiple disciplines involved in anomaly detection. We strived for obtaining solid knowledge on this topic. Our published work [7] summarizes the outcome of our survey. We extended the contents of that paper in Chapter 1 of this thesis.  We also learned practical issues about data pre-processing to train a neural artificial network while we worked on another published paper that concludes the previous research.

- ▪ Data Verification

  This phase involved the quality assessment of the data provided. Two different problems were considered. One of them was documented in a report to the capture machine manufacturer and raised deep investigation on their side. More details are given on the next Chapter.

- Data Preparation

  The preparation of 6.5 TB of raw binary data packets for our analysis was done in several steps. To illustrate, a capture representing 10 minutes of network traffic (at early hours of the night) had a rate of 193,556 packets/sec with an average packet size of 1,066 bytes. When using Bro, this input generates approximately 1.6 millions of connections with more than 88 thousands of anomalous events. Conversely, Snort raised 5.8 millions of alerts (the alerts include a large amount of anomalous events indicating fragmentation observed by Snort pre-processors). This numbers decreases oscillating as network load diminishes over the night. Later in this chapter, we give details on the setup configuration for both systems. Throughout the thesis, the other considerations are documented.

- Data Analysis

  In this thesis, we fed Bro with data representing 500 minutes of traffic. We also verified some cases with Snort; data input involves 20% of the total traffic. The outputs were overwhelming. We developed ad-hoc scripts, load events in databases and generate plots to facilitate the data analysis. The next section explains why and how we did it.

## 2.2 Data Analysis

### 2.2.1 Overall Approach

Security monitors, e.g. IDS, analyze traffic and record anomalous activity and raise alarms when an attack is recognized. Anomalous events are warning us that something may be wrong; they do not necessary indicate an attack. We believe that the study of anomalies is important. Proliferating botnets act surreptitiously. Behind botnets there is an underground economy that impulse the development of sophisticated mechanisms to abuse our networks without being discovered. These mechanisms might be detected by the verification of anomalous behavior.

A caveat is that anomaly activity increases the uncertainty in the analysis. The number of anomalies events surpassed the number of alerts triggered by sensors; it is more difficult to understand the causes; and a considerable part may not indicate malicious activity. Usually, security administrators overlook anomalies.

Machine learning techniques could be a solution to solving some of the problems above mentioned. To make use of this approach, first we need to understand better the nature of anomalies and evaluate them in context with the environment. Traffic network analysis is a must.

Considering our network environment, there is neither the previous statistics on the traffic network behavior at the main gateway, nor even the estimation of the basic network characteristics. As a result, we would like to have mere estimation on the number of hosts inside U-Tokyo network. Currently, there is no answer to our question.

Consequently, prior to analyze anomalies, we examine the traffic behavior and attempt to profile the major characteristics.

A limitation is that we are not able to contrast our statistics with the previous metrics. However, we believe that in this context our work may shed a brighter light on what we know about U-Tokyo network and bring attention to a topic that was not considered in the previous research. Combination of network measurement study and security study will open more opportunities for further research.

### 2.2.2 Anomalies Observation Method

Anomalies are detected using intrusion detection systems. Bro IDS is the main system and Snort IDS is the second monitor. We conduct analysis that covers a wide aspect of anomalous behavior in context with the environment where the anomalies are observed. To achieve this goal we analyze 35 millions (M) of incoming connections and 26 M of outgoing connections to observe patterns in basic features such as connections, traffic amount; in aggregate data to analyze frequent episodes to illustrate some Dos and Probing attacks by observing SYN and fragments among other indicators. Then, we scrutinize activity not considered in the KDD dataset using indicators such as more than 1 M of anomalies, host identified as not following standard and evasion techniques that might be associated with botnets. Finally, we conclude with a global view of the anomalies found.

The minimal checklist for assessment the anomaly includes:

- Generate plots discriminating incoming and outgoing activity for each network
- Detect interesting patterns
- Check source hosts (e.g. distinct host, distinct subnet class C, top-ranked host)
- Check destination hosts

- Check source port
- Check destination port
- Check connections flag status and services/protocols involved (anomaly in context)
- Check other anomalies and alerts triggered by the top-ranked host (anomaly in context)

## 2.3 Tools

Security network research is focused, roughly speaking, on protecting resources from threats that originate either outside or inside of an organization. To pursue this goal, an intrusion detection model was proposed in 1987 [26] and many Intrusion Detection Systems (IDS) were developed based on that model. The taxonomy defined in [27] provides a survey and classification of different IDS approaches to understand the principle of operation, underlying mechanisms, the targeted intrusion and environment that govern the detection strategy. A more recent wide-ranging survey of intrusion detection efforts is in [28]. For an extensive review of computational intelligence (CI) methods that includes artificial neural networks, fuzzy sets, evolutionary computation, artificial immune systems, swarm intelligence and soft computing, see [29].

Two major approaches govern the research on intrusion detection: misuse detection and anomaly detection. However, it is generally difficult to label a state-of-the-art solution using only one of these categories. On one side, misuse detection systems are typically based in signatures to recognize known attacks. They are widely accepted and deployed in industrial environments because its good performance in high-speed networks. The need for frequent maintenance to avoid becoming outdated against the evolving nature of threats is a major drawback. As an example, the number of new and updates rule in the VTR ruleset for Snort version 8.6.1 in the last 180 days was 214 and 157[1] respectively. On the other side, anomaly detection systems use a profile for normal and legitimate traffic to establish a baseline and look for deviations from this normal profile. They may be helpful in recognizing unknown attacks. The two paradigms are complementary, and some solutions integrate both of them to reinforce countermeasures.

---

[1] Number was calculated using the change-log pages between 13-Ago-10 to 12-Jan-11. Source: from the Sourcefire website .We could not find updates in Nov and updates of 27-Sep-10 and 12-Oct-10, which contains the major numbers of changes, are obtained from a cached.

### 2.3.1 Bro IDS

This section summarizes information obtained in [30]. Bro IDS is a network-based intrusion detection system that monitors passively the traffic seen on a network link. Its core is an event engine that pieces network packets into events that reflect different type of activity. Events are actions that take place on the network. Examples of events might be a failed connection attempt, a TCP packet with a bad checksum, a successful connection or an IRC invalid command.

**Philosophy**

It was originally designed by Vern Paxson of ICSI' s Center for Internet Research (ICIR), Berkeley to provide an open-source research framework for online network analysis. Project started in 1995 at Lawrence Berkeley National Laboratory (LBL) and Bro was first published in 1998. Its design's concepts put emphasis on the application-level semantics and tracking information over time. Pre-written policy scripts are mostly neutral, i.e. no presumption of "good" or "bad". An analyst can use a specialized Bro language to write domain-specific policy scripts.

**Arquitecture**

Illustrated in Figure 2.1

**Analysis model**

Analysis model is not signature-matching, though Bro provides support for byte-stream/packet-based signature matching, including context. In the past, the python script snort2bro was used to translate Snort signatures to Bro language but today Snort newer options difficult an automatic translation. Analysis model is not anomaly detection, though Bro is usable for that in principle.

**Logs**

Currently, the event engine processes more than 300 events that are handled by analyzers. Bro provides analyzers to cover different aspects of the connection, i.e. generic connection analysis at the level of TCP or UDP and specific analysis for an application on top of TCP.

Usually, detection is expressed in terms of "notice" and eventually is promoted to "alarm".

Figure 2.1 – Bro Arquitecture [30]

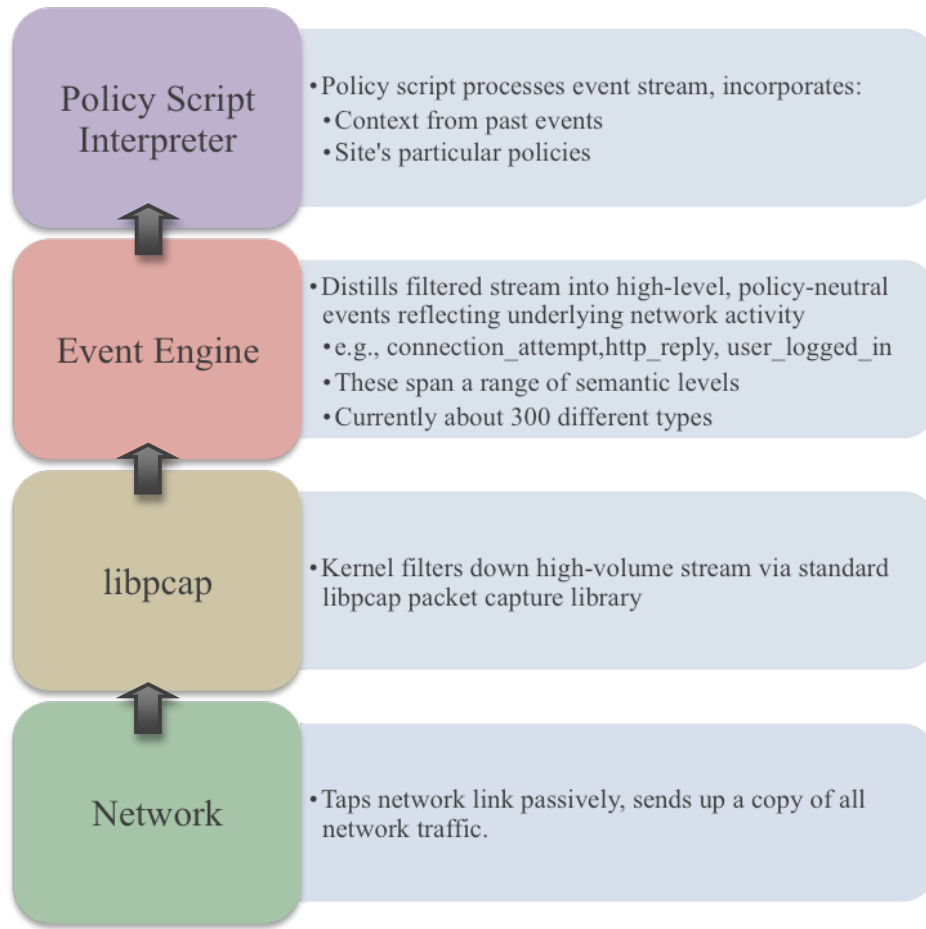

Each analyzer logs its output in ASCII files. The main files are:

- **conn.log**

One line represents a summary of each connection seen by Bro, the format is shown below:

```
<start> <duration> <local IP> <remote IP> <service> <local port> \
<remote port> <protocol> <org bytes sent> <res bytes sent> <state> \
<flags> <tag>
```

An example of a connection summary:

```
931803523.006848    54.3776    http    7320    38891    206.132.179.35
128.32.162.134 RSTO X %103
```

The connection began at Unix time 931803523.006848 (18:18:43 hours GMT on July 12, 1999) and lasted 54.3776 seconds. The service was HTTP (presumably). The originator sent 7,320 bytes, and the responder sent 38,891 bytes. Because the L flag is absent, the connection was initiated by host 128.32.162.134, and the responding host was 206.132.179.35. When the summary was written, the connection was in the RSTO state.

There is additional information, summarized by the string %103 logged in the http log.

- **weird.log**

Weird analyzer processes unusual or exceptional events. Such events can reflect incorrect assumptions, attempts by attackers to confuse the monitor and evade detection, broken hardware, misconfigured networks, and so on. We use this file as a source of anomalies.

For each event, it is possible to configure the action taken by the analyzer. An event could be ignored, recorded each time it occurs, recorded the first time it occurs for a given connection or the first time it occurs for a given originating host.

- **notice.log**

It is the primary output facility. Distributed scripts detect 78 different types of notices. Notices do not necessarily convey a problem but scripts provide number of ways to decide when to promote a notice to an "alarm".

- **alarm.log**

Alarms can be triggered when network traffic matches signatures and when matches rules embedded in analyzers. Notices and weird events can be promoted to alarms.

**Dynamic Detection Protocol (DDP)**

When the DDP framework is loaded, every connection is associated with an analyzer tree that contains an arbitrary number of analyzers that can be modified during the lifetime of the connection. Old versions of Bro used the port number to launch an analyzer but, with the new capability, Bro can use the payload to make a decision about the correct analyzer independently of the port number. Analyzers are turned off dynamically when they are detected as wrong analyzer [31].

**Selection Criteria**

The aforementioned characteristics make Bro a unique platform, well aligned for behavioral analysis, policy enforcement and specification-based intrusion detection.

We use Bro as the main system to characterize and analyze anomalies.

### 2.3.2   Snort IDS

This section summarizes information obtained in the Snort User Manual and README files included in Snort code, which can be found in [32]. Snort is a network-based intrusion detection system that can be used as a straight packet sniffer like tcpdump, a packet logger or as a network intrusion prevention system. Its core is the detection engine and the rules database.

Snort performs protocol analysis, content searching/matching, and is commonly used to actively block or passively detect a variety of attacks and probes, such as buffer overflows, stealth port scans, web application attacks, SMB probes, and OS fingerprinting attempts, amongst other features.

**Philosophy**

Released originally by Martin Roesch in 1998, developed and maintain by Sourcefire (a company founded by the same author). The Snort project relies on open source development methodology. People in the open source Snort community worldwide can detect and respond to bugs and other security threats.

Sourcefire offers a commercial version of the Snort technology that includes an easy-to-use interface, optimized hardware, powerful data analysis and reporting, policy management and administration, a full suite of product services, and 24x7 support.

Subscribers receive new and updates to rules in real time, registered users have access to the same rules 30 days after release.

**Arquitecture**

First, traffic is acquired from the network link via libpcap. Packets are passed through a series of decoder routines that first fill out the packet structure for link level protocols then are further decoded for things like TCP and UDP ports.

Packets are then sent through the registered set of preprocessors. Each preprocessor checks to see if this packet is something it should look at.

Packets are then sent through the detection engine. The detection engine checks each packet against the various options listed in the Snort rules files. Each of the keyword options is a plugin; then, is easily extensible.

**Analysis Model**

Analysis model is mainly based on rules, which are more powerful than signatures. According to the definition given by Sourcefire a signature-matching approach is used for matching a particular exploit or unique pattern, conversely a rule checks vulnerabilities. It is necessary to understand the nature of the vulnerability to develop the rule.

Classification of the rule set:

- Not suspicious traffic
- Unknown traffic
- Potencially bad traffic
- Attempted Information Link
- Attempted Denial of Service
- Attempted User Privilege Gain
- Unsuccessful User Privilege Gain
- Attempted Administrator Privilege Gain
- Successful Administrator Privilege Gain

**Preprocessors**

Preprocessors are a key part of the Snort arquitecture. They are called before the detection engine. There are many preprocessors and users can extend them. We introduce the two preprocessors related to our analysis. Both are target-based, i.e. the system tries to recreate the behavior of the target system. We will talk more about this capability in the Section 4.2.1.

Frag3: IP defragmentation module. It can alerts 13 anomalies:

1. IP Options on fragmented packet
2. Teardrop attack
3. Short fragment, possible DoS attempt
4. Fragment packet ends after defragmented packet
5. Zero-byte fragment
6. Bad fragment size, packet size is negative
7. Bad fragment size, packet size is greater than 65536
8. Fragmentation overlap
9. IPv6 BSD mbufs remote kernel buffer overflow
10. Bogus fragmentation packet. Possible BSD attack
11. TTL value less than configured minimum, not using for reassembly

12    Number of overlapping fragments exceed configured limit

13    Fragments smaller than configured min_fragment_length

Stream5 is a TCP reassembly module with ability to track TCP and UDP sessions. TCP sessions are identified using the 3-way handshake. UDP sessions are established using consecutive series of UDP packets from two end points via the same set of ports. It also tracks ICMP for checking messages that will end TCP and UDP connections (unreachable and service unavailable messages). It can alerts 10 TCP anomalies:

1    SYN on established session

2    Data on SYN packet

3    Data sent on stream not accepting data

4    TCP Timestamp is outside of PAWS window

5    Bad segment, overlap adjusted size less than/equal 0

6    Window size (after scaling) larger than policy allows

7    Limit on number of overlapping TCP packets reached

8    Data after Reset packet

9    Possible Hijacked Client

10   Possible Hijacked Server

**Logs**

The way Snort logs events is user-configurable and allows several methods. We opted for logging events directly in a database. This is not the fastest and efficient option, but considering that our analysis was offline, our priority was the convenience of realizing SQL queries on the event tables.

**Selection Criteria**

According to Sourcefire website, Snort is the world's most downloaded IDS with nearly 4 millions of downloads to date and approximately 300.000 registered users. Because it can be download for free, the number of downloads does not indicate the number of Snort installations, however, it is a good indicator of its popularity.

## 2.4    Other considerations

### 2.4.1    Argus

When considering the adequate tool for helping with the traffic analysis, we evaluated Argus [33]. This software is a Network Audit Record Generation and Utilization System. It processes packets and generates detailed status reports of the flows detected in the packet stream. The user has a wide variety of options to aggregate traffic and define flows according the needs of the project. Several published works are based on Argus [34, 35, 36].

We conducted basic experiences using Argus on part of the capture, but the output flows differed from Bro output. It was not easy to associate some anomalies events triggered by Bro with the flows recognize with Argus. Differences in the number of IP addresses detected by both systems made us desist of using Argus to reduce sources of uncertainties. However, we consider our tests as preliminary work and this subject deserves more attention in future research.

### 2.4.2   System Configuration

Bro is applied to the totality of data and Snort to a 20% of data.

**Bro**

To execute Bro we run this command on 50 subsets of captured data, called "periods":

> bro *$files site-policy* –f *tcpdump-filters variables list-of-analyzers*

**$files:**

This variable contains a list of consecutive files that in total represents 10 minutes of network traffic (1 period)

**Site policy:**

We indicate the 3 main blocks of IP class B to be considered as our local networks.

 @load site
redef local_nets += { x1.x2.00/16, y1.y2.0.0/16, z1.z2.0.0/16 };

For compatibility with Snort, we specify:
redef tcp_reassembler_ports_orig: set[port]+=
{21/tcp,22/tcp,23/tcp,25/tcp,42/tcp,53/tcp,79/tcp,109/tcp,110/tcp,111/tcp,113/tcp,119
/tcp,135/tcp,136/tcp,137/tcp,139/tcp,143/tcp,161/tcp,445/tcp,513/tcp,514/tcp,587/tcp
,593/tcp,691/tcp,1433/tcp,1521/tcp,2100/tcp,3306/tcp,6665/tcp,6666/tcp,6667/tcp,66
68/tcp,6669/tcp,7000/tcp,32770/tcp,32771/tcp,32772/tcp,32773/tcp,32774/tcp,32775
/tcp,32776/tcp,32777/tcp,32778/tcp,32779/tcp,80/tcp,443/tcp,465/tcp,563/tcp,636/tc
p,989/tcp,992/tcp,993/tcp,994/tcp,995/tcp,1220/tcp,2301/tcp,3128/tcp,6907/tcp,7702
/tcp,7777/tcp,7779/tcp,7801/tcp,7900/tcp,7901/tcp,7902/tcp,7903/tcp,7904/tcp,7905/

tcp,7906/tcp,7908/tcp,7909/tcp,7910/tcp,7911/tcp,7912/tcp,7913/tcp,7914/tcp,7915/t
cp,7916/tcp,7917/tcp,7918/tcp,7919/tcp,7920/tcp,8000/tcp,8008/tcp,8028/tcp,8080/tc
p,8180/tcp,8888/tcp,9999/tcp};

redef tcp_reassembler_ports_resp: set[port]+=
{80/tcp,443/tcp,465/tcp,563/tcp,636/tcp,989/tcp,992/tcp,993/tcp,994/tcp,995/tcp,122
0/tcp,2301/tcp,3128/tcp,6907/tcp,7702/tcp,7777/tcp,7779/tcp,7801/tcp,7900/tcp,790
1/tcp,7902/tcp,7903/tcp,7904/tcp,7905/tcp,7906/tcp,7908/tcp,7909/tcp,7910/tcp,791
1/tcp,7912/tcp,7913/tcp,7914/tcp,7915/tcp,7916/tcp,7917/tcp,7918/tcp,7919/tcp,792
0/tcp,8000/tcp,8008/tcp,8028/tcp,8080/tcp,8180/tcp,8888/tcp,9999/tcp};

**tcpdump filters:**

"tcp or udp or icmp or (ip[6:2] & 0x3fff != 0)"

**variables:**

dpd_conn_logs=T

**list-of-analyzers**:

dpd detect-protocols dyn-disable detect-protocols-http proxy ssh irc-bot brolite print-globals capture-loss

### Snort

All the system configuration is established in the snort.conf file:

**# Setup the network addresses you are protecting**
var HOME_NET [x1.x2.00/16, y1.y2.0.0/16, z1.z2.0.0/16]

**#Servers on your network**

$HOME_NET

**# Target-based IP defragmentation.**

preprocessor frag3_global: max_frags 1048576 , memcap 1073741824

preprocessor frag3_engine: policy windows detect_anomalies overlap_limit 10
min_fragment_length 100 timeout 180

**# Target-Based stateful inspection/stream reassembly**.

preprocessor stream5_global: max_tcp 1048576, max_udp 1048576, memcap
1073741824, track_tcp yes, track_udp yes, track_icmp no

preprocessor stream5_tcp: policy windows, detect_anomalies, max_queued_segs
10485760, require_3whs 180, \

### 2.4.3 Future Versions

We use for the analysis Bro Version 1.5.1 and Snort Version 2.8.6.1 (using PCRE version: 8.10 2010-06-25 and ZLIB version: 1.2.3.3). Snort rule set corresponds to VTR rules of Jun 23th, 2010.

Both systems are planning major changes. Bro mailing list communicated that Bro Project received a 3 year grant from the National Science Foundation for improving the project on all fronts including code quality and supportability, "out of the box" detections, improved user support, and new documentation. A new version of the software is planned to be release this year. Snort Version 3 was release more than 2 years ago, but it is still in beta. According to Snort designer "the new version 3 will serve as a network traffic analysis platform as well. The arquitecture is less susceptible to IDS/IPS evasion or bypass attacks, where attackers sneak past the devices."

# Chapter 3

# Traffic Characteristics

## 3.1   Collected Data

Full payload packets were collected at the main campus link of University of Tokyo (abbreviated Todai) during night hours from July 15th 2009 6:53 PM to July 16th 2009 6:43 AM. In this study, we present the analysis on 50 selected periods; each representing 10 minutes of captured traffic. The total amount of analyzed packet captures is 4.7 TB split in pcap files of 4 GB. Network activity decreases as time progresses. First file capture at 7:24 PM represents 21 seconds of traffic at a rate of 186,673.24 packets/sec and last file capture 6:13 AM represents 72 seconds of traffic at a rate of 62,112.21 packets/sec. Time span between periods is indicated in Figure 3.1. We will refer to the total analyzed traffic as the Todai trace.

The results of the statistical analysis are presented for the three main blocks of IP address class B discriminated in outbound and inbound connections. Unless otherwise specified, the term "Net" plus a number suffix, e.g. Net-1, represents a network class B and the term "subnet" represents a subnet class C. Numbers are used consistently to facilitate reading and comparison.

Traffic characteristics were measured using different systems and tools: Bro IDS, Argus, capinfos, trace-summary and ad-hoc scripts. We mainly provide results obtained through the analysis of connection summary logs from Bro to show coherent and meaningful data that can be contrasted to the anomalies observed by the same system. Then, the unit of analysis is a connection.

Figure 3.1 – Gap of Time Between Selected Periods. x-axis: Period, y-axis: Seconds

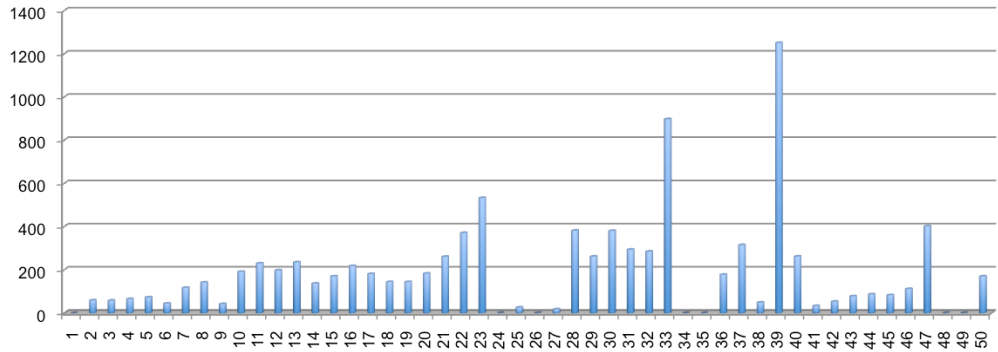

To conduct this study, the only information we have about network usage is that Net-2 is the main block and together with Net-1 the address usage is high. Net-3 is an old block with moderate usage.

Regarding the number of host using U-Tokyo network, we have only the IP address as a reference (both IDS record only IP addresses to identify hosts). DHCP and NAT servers make difficult to estimate the actual number. We consider, with strong likelihood, that during a 10 minutes period, the number of IP addresses observed in the traffic equals the numbers of host. In Todai trace, that quantity oscillates between 3200 and 7400 hosts with exceptions during bursty traffic.

## 3.2   Limitations in the Capture Process

The quality of the data is a critical factor to assure reliability of our analysis. The quality assessment was a difficult task since we did not participate in the capture process. However, we noticed two deviations that are documented in this section.

### 3.2.1   Malformed Packets

The first problem was difficult to detect. Malformed packets caused by the capture process introduced unpredictable results. However, the anomaly was hidden in the merging process of multiple small files and when was unveil it was confused with a traffic anomaly.

The original capture comprise 1632 pcap files of 4 GB, each representing approximately 20 seconds of traffic. Bro IDS does not memorize internal states for connections that cross file boundaries. Thus, it is recommendable to merge small files in a large pcap file for offline analysis. Another approach is to simulate an online analysis by feeding the pcap files into the Click! Modular Router [37]. Click! Router is able to load balance the traffic to feed a Bro cluster with many workers to leverage multiple cores and speed up the analysis process.

We opted for the merging solution using the tool ipsumdump developed by the same author of the Click! Router. Files were merged previous to feed Bro IDS without raising any error or warnings. Then, while checking closely a particular attack, a worrisome message triggered by the network protocol analyzer (Wireshark) thwarted all attempts of reading the pcap file. An investigation showed that 32 files have the same pattern. Different tools processed the malformed packets inconsistently. Capture machine developer engineers are now studying the problem.

The limitation with the capture machine forces us to analyze data in ten minutes periods. Experiments are conducted using 50 periods within 11 hours of captured traffic. The interval between periods is 3 minutes on average with the exception of gap between periods 32-33 and 38-39. Details are indicated in Figure 3.1.

### 3.2.2    Noncompliance 3-way Handshake

The second potential problem was revealed in other experiments conducted on the laboratory at the time of writing this thesis.

The transmission in TCP Protocol is connection-oriented. Before any data can be transmitted, a reliable connection must be obtained and acknowledge. The process to establish the connection is known as the initial 3-way handshake. This initial communication assures that client and server synchronize each other's sequence numbers.

A normal 3-way handshake starts when the client sends a SYN containing the segment's sequence number to the server. The server responds with a SYN-ACK to acknowledge the client's sequence number and also to transmit its sequence number. The connection is correctly established when the client sends and ACK to the server.

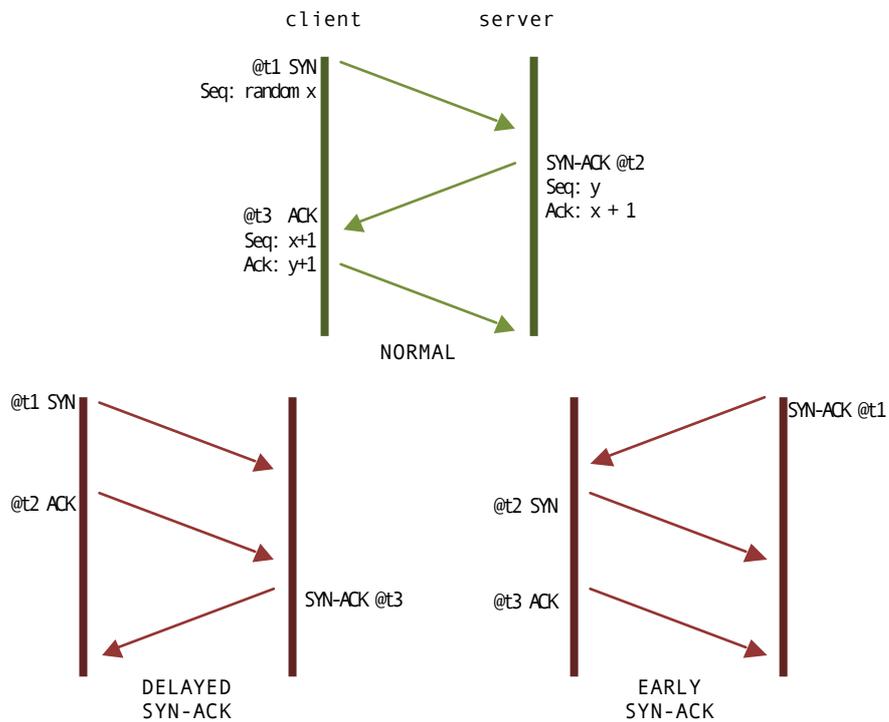

Figure 3.2 – Artifacts in Initial 3-way Handshake.
Above: normal, below: delayed and early syn-ack

We observe some exceptional handshakes where SYN-ACK was transmitted before the SYN, "early SYN-ACK", and some SYN-ACK appear after the ACK, "delayed SYN-ACK"; problem is illustrated in Figure 3.2. If this irregularity is an artifact introduced by the capture machine (or by the mirror port of the router used for the capture), it may cause trouble to Bro IDS, which assumes that it sees packets in the order they were in the wire. In that case, the outcome of our analysis would be affected. A possible explanation would be an out of synchronization between the uplink and the downlink wire causing differences in the timestamp of packets coming from different directions.

To validate our data, we look closer at the initial handshake process initiated by more than 28,000 SYN and 20,000 SYN-ACK observed in 4 GB of traffic (one pcap file). An initial thought was to consider that fast SYN – SYN-ACK and SYN-ACK – ACK answers might be affected. Our metrics show the time span between the arrival of SYN and the associated SYN-ACK (SYN<>SYN-ACK), between the SYN-ACK and ACK (SYN-ACK<>ACK), and between SYN and ACK (SYN<>ACK) for three possible cases: normal, early SYN-ACK and delayed SYN-ACK in inbound and outbound connections. To simplify the problem we did not consider retransmissions to calculate average times, instead we used the last SYN observed and the first ACK observed.

Table 3.1 – Metrics on Initial 3-way Handshake.

Unit: μsec – (standard deviation between brackets)

| Early SYN-ACK | | | | Delayed SYN-ACK | | | |
|---|---|---|---|---|---|---|---|
| #33 cases in connections initiated by Todai #20 cases in connections initiated remotely | | | | #29 cases in connections initiated by Todai #9 cases in connections initiated remotely | | | |
| SYN <> SYN-ACK | | SYN <> ACK | | SYN <> ACK | | SYN-ACK <> ACK | |
| TODAI | REMOTE | TODAI | REMOTE | TODAI | REMOTE | TODAI | REMOTE |
| 36039 (171233) | 28034 (90057) | 5221 (12044) | 215488 (214601) | 356800 (839634) | 2028927 (2113325) | 798738 (1620489) | 2370391 (1608777) |
| Normal Handshake | | | | | | | |
| # 14,046 cases in connections initiated by Todai #5,686 cases in connections initiated remotely | | | | | | | |
| SYN <> SYN-ACK | | SYN <> ACK | | | | SYN-ACK <> ACK | |
| TODAI | REMOTE | TODAI | | REMOTE | | TODAI | REMOTE |
| 62617 (142540) | 19792 (244187) | 69296 (161855) | | 243318 (743532) | | 6678 (75662) | 223525 (670208) |

However, ambiguities in retransmissions make these calculations more complicated [38]. Results are presented in Table 3.1; standard deviation is between brackets.

Normal handshake duration of connections locally initiated is obviously shorter since the capture machine, which is located at the U-Tokyo gateway, receives the SYN and ACK faster than connections externally initiated. The handshake process of early SYN-ACK cases, especially in connections generated locally is considerable shorter than normal connections. Though, the difference was generated because of our simplification of the problem. When checking manually the abnormal handshake situations were generated in cases of early SYN retransmissions and early SYN-ACK retransmissions.

In our pcap file, a high percentage of 26% of the SYN and 8% of SYN-ACK are retransmissions. SYN retransmissions in benign traffic are expected to happen because at the moment the SYN is sent the round-trip delay has not been yet calculated and estimates that are too low generate SYN retransmissions [39].

Attacks may also explain retransmissions. In SYN flooding, an attacker generates a huge amount of half-open connections to exhaust server resources. The server when it first receives the SYN, answers immediately with a SYN-ACK. This attack is usually combined with IP spoofing. Thus, the ACK packet will never arrive. However, the unaware server will retransmit SYN-ACK a number of times to give the client a chance to resend the ACK packet. We have observed a slow amount of partial connections over the 20 seconds studied on this section. However, this type of attack is often seen in Todai trace and presented in Section 3.4.

Table 3.2 – Detections using DPD / Inbound Connections

| | Detected non-std. port | | Rejected by analyzer non-std. port | | Rejected by analyzer std. port | |
|---|---|---|---|---|---|---|
| | # conn | # local IP | # conn | # local IP | # conn | # local IP |
| **Net-1** | | | | | | |
| HTTP | 301 | 31 | 120 | 2 | 2620 | 129 |
| FTP | 0 | 0 | 0 | 0 | 0 | 0 |
| IRC | 1 | 1 | 0 | 0 | 0 | 0 |
| SMTP | 0 | 0 | 0 | 0 | 0 | 0 |
| **Net-2** | | | | | | |
| HTTP | 7376 | 29 | 44587 | 4 | 9096 | 821 |
| FTP | 3 | 1 | 0 | 0 | 0 | 0 |
| IRC | 1 | 1 | 2 | 2 | 0 | 0 |
| SMTP | 0 | 0 | 0 | 0 | 17 | 3 |
| **Net-3** | | | | | | |
| HTTP | 67 | 9 | 0 | 0 | 1030 | 17 |
| FTP | 0 | 0 | 0 | 0 | 0 | 0 |
| IRC | 0 | 0 | 0 | 0 | 0 | 0 |
| SMTP | 0 | 0 | 0 | 0 | 0 | 0 |

## 3.3 Traffic classification

The feature Dynamic Detection Protocol (DPD) [31] is used for traffic classification. We reproduced the metrics of Table 5 published in the aforementioned paper on Table 3.2 and Table 3.3. Unlike the original table, in this study we want to highlight the differences among inbound and outbound traffic, and also the number of local IP addresses involved in the activity. For our purposes, the merely indication of connections may give an erroneous impression of the magnitude of the problem. By checking local IP addresses we observed that thousand of connections triggering "Protocol Violations" are reduced in some cases to only 4 local IP addresses.

Compared to the table published in 2006, we observe less activity of FTP, IRC or SMTP connections.

Table 3.3 – Detections using DPD / Outbound Connections

| | Detected and verified non-std. port | | Rejected by analyzer non-std. port | | Rejected by analyzer std. port | |
|---|---|---|---|---|---|---|
| | # conn | # local IP | # conn | # local IP | # conn | # local IP |

Net-1

| | | | | | | |
|---|---|---|---|---|---|---|
| HTTP | 65821 | 1031 | 144 | 20 | 12871 | 450 |
| FTP | 29 | 23 | 0 | 0 | 0 | 0 |
| IRC | 561 | 15 | 0 | 0 | 0 | 0 |
| SMTP | 3 | 3 | 0 | 0 | 100 | 24 |

Net-2

| | | | | | | |
|---|---|---|---|---|---|---|
| HTTP | 65821 | 1031 | 7293 | 31 | 10228 | 567 |
| FTP | 29 | 23 | 0 | 0 | 1909 | 2 |
| IRC | 561 | 15 | 2 | 1 | 64 | 3 |
| SMTP | 3 | 3 | 0 | 0 | 424 | 33 |

Net-3

| | | | | | | |
|---|---|---|---|---|---|---|
| HTTP | 65821 | 1031 | 13 | 4 | 431 | 46 |
| FTP | 29 | 23 | 0 | 0 | 0 | 0 |
| IRC | 561 | 15 | 0 | 0 | 0 | 0 |
| SMTP | 3 | 3 | 0 | 0 | 21 | 7 |

The first column shows how often a protocol detection signature flagged the given protocol as running on a non-standard port, for which the corresponding analyzer verified the detection. In the case of outbound connections we cannot discriminate which local net initiated them. Bro DPD logs only the server IP address, for further research connection logs should be checked.

Most prominent HTTP inbound connections on non-standard ports are due to web application servers, peer to peer web content distribution, proxy servers, file sharing services and security agents. Most popular applications in Net-1 are Apache, SSH, Delegate, Mongrel, squid, FileMakerPro, CommuniGatePro, Trend. In Net-2, applications found are Apache, ligthttpd, nginx, Microsoft-IIS, Google Web Server, BWS, SSH, Amazon S3 and Coral CDN proxy. In Net-3, the common applications found are Zope, Apache, lighttpd, and squid.

The second and third columns list how often the analyzer did not agree with the detection, but instead rejected the connection as exhibiting the given protocol, for non-standard and standard ports, respectively. We have manually inspected a subset of protocol failures: SMTP failures are due to reply codes beyond the assigned range, IRC failures are mostly for extremely short messages and invalid reply number, 63% of the total HTTP protocol violations are requests that does not follow the standard and the rest are replies showing the same behavior, finally FTP generates messages of non numeric reply code.

## 3.4    Main Services and Applications

### 3.4.1    Inbound and Outbound Connections

According to Bro Reference Manual:

- TCP protocol clearly defines the establishment and termination of TCP connections. Bro uses the variable *tcp_inactivity_timeout* variable for posterior connection classification in different states.
- For UDP, a connection begins when host *A* sends a packet to host *B* for the first time, *B* never having sent anything to *A*. This transmission is termed a *request*, even if in fact the application protocol being used is not based on requests and replies. If *B* sends a packet back, then that packet is termed a *reply*. Each packet *A* or *B* sends is another request or reply. Bro UDP connection timeouts are specified through the *udp_inactivity_timeout* variable.
- For ICMP, Bro likewise creates a connection the first time it sees an ICMP packet from *A* to *B*, even if *B* previously sent a packet to *A*, because that earlier packet would have been for a different *transport* connection than the ICMP itself---the ICMP will likely *refer* to that connection, but it itself is not part of the connection. Bro ICMP connection timeouts are specified through the *icmp_inactivity_timeout* variable.

Graphs for each net with the distribution of incoming and outgoing connections for the top 15 network services (application breakdown) are presented from Figure 3.3 to Figure 3.8. The totality of connections registered in *conn.log* file is used. Timeouts for connection definition are not modified; the default values are specified in *Bro.init* file.

Traffic was captured during the less busy hours of the network; then, some activities gradually diminish as hour progress. Other activities remain almost constant. Bursty incoming traffic from particular services sometimes duplicates the amount of connections found in the affected period.

The ratio of incoming and outgoing connections for both Net-1 and Net-2 is 1:1 while in Net-3 the ratio is 5:1. We have processed approximately 10 million (10M), 14M and 11M of incoming connections for each network respectively. Another observation is the difference in the connection transport protocol. In Net-1, both incoming and outgoing connections present a roughly proportion of 65% TCP connections and 35% UDP connections. Net-2 is characterized with more presence of ICMP (3%) for both types of connections, less presence of TCP in outgoing connections (55%) but more presence in incoming connections (76%). Finally, Net-3 outgoing connections have a significant lower presence of TCP protocol (39%) that might be explained with the decrease in HTTP outgoing activity.

**Connection states**

States reflect the connection status at the time the Bro *conn.log* file is written, which is usually when the connection terminates or Bro terminates. The system uses 13 different flags to indicate:

- S0  Connection attempt seen, no reply.
- S1   Connection is established, not terminated.
- SF  Normal established and termination.
- REJ   Connection attempt rejected.
- S2  Connection established and close attempt by originator seen.
- S3  Connection established and close attempt by responder seen.
- RSTO  Connection established, originator aborted (sent a RST).
- RSTR  Established, responder aborted.
- RSTOS0  Originator sent a SYN-ACK followed by a RST.
- RSTRH Responder sent a SYN-ACK followed by a RST.
- SH   Originator sent a SYN followed by a FIN.
- SHR  Responder sent a SYN-ACK followed by a FIN.
- OTH  No SYN seen, just midstream traffic (a "partial connection")

In Todai trace, connections are usually labeled SF or S0 but each of the analyzed networks is characterized with different proportions. According to the robustness principle of TCP protocol *"be conservative in what you do, be liberal in what you accept from others"* outgoing connections are expected to be flagged as normal established and terminated. Although an important part of the connections are flagged S0, OTH or RSTO, a conservative rule applies to the outgoing traffic observed in U-Tokyo. Net-3 outgoing connections are the more conservative in the sense of that principle, with a 67.9% of connections flagged "SF". Incoming connections have a lower percentage of SF connections, and a high percentage of connections labeled S0.

Another significant aspect to consider is the transmitted bytes. A different application breakdown emerges by using this variable; results are presented in the next section.

Figure 3.3 – Net-1: Inbound Connections

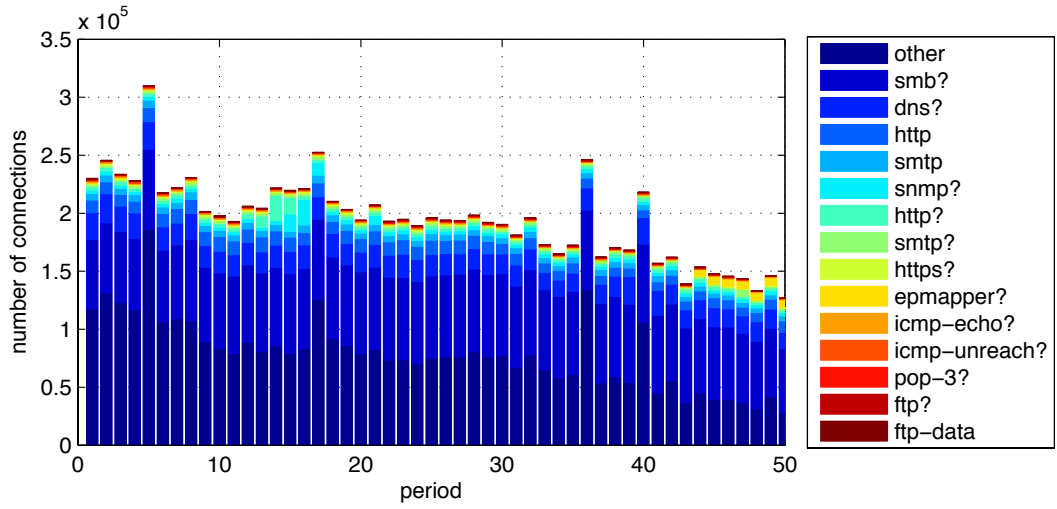

Figure 3.4 – Net-1: Outbound Connections

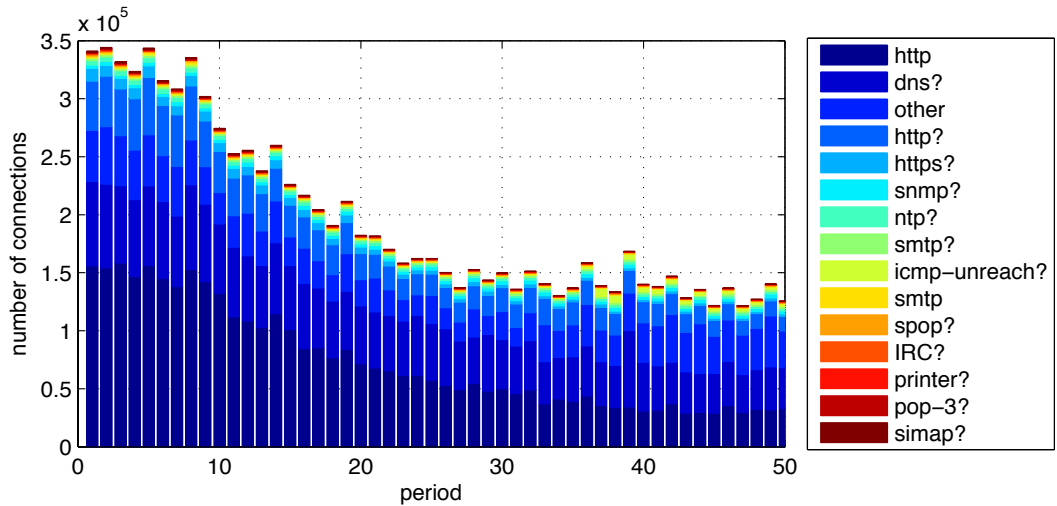

Figure 3.5 – Net-2: Inbound Connections

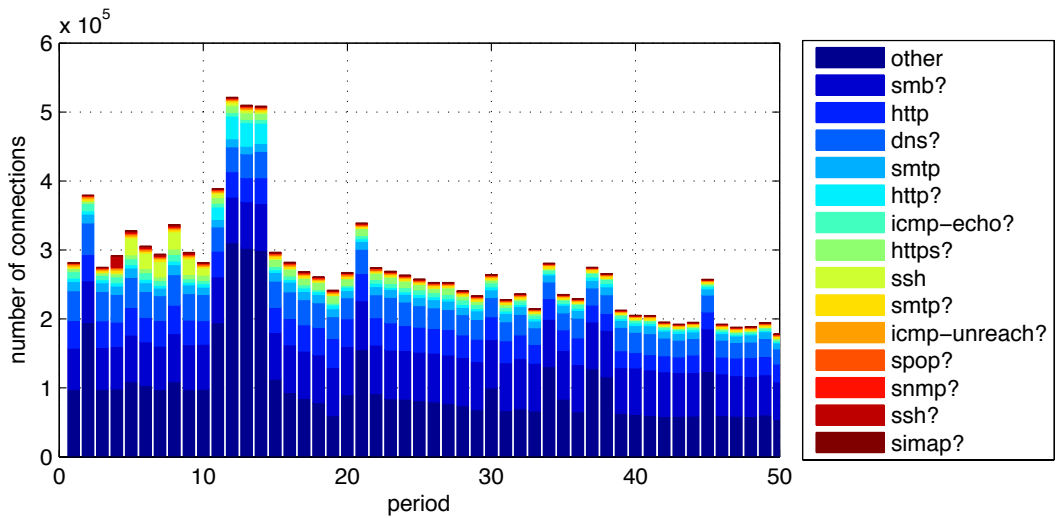

Figure 3.6 – Net-2: Outbound Connections

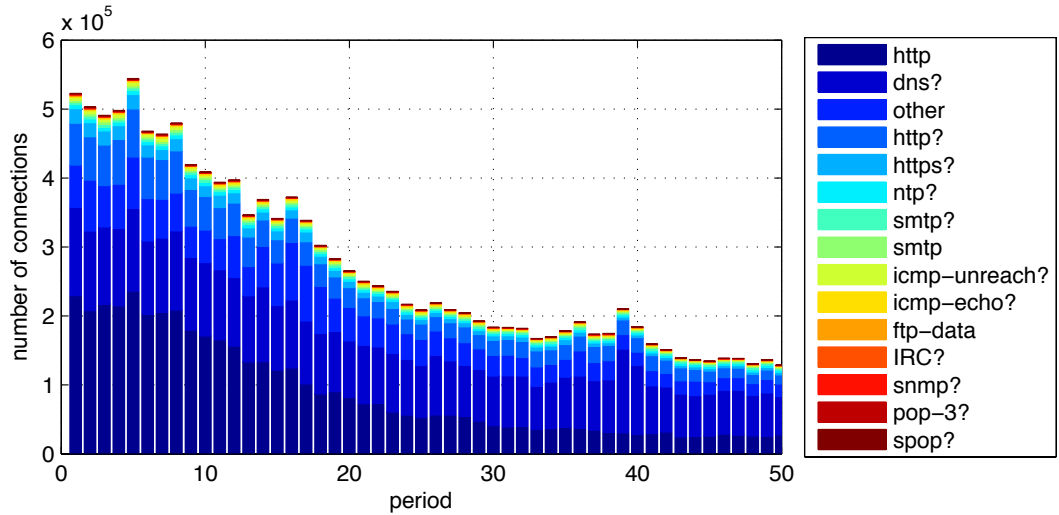

Figure 3.7 – Net-3: Inbound Connections

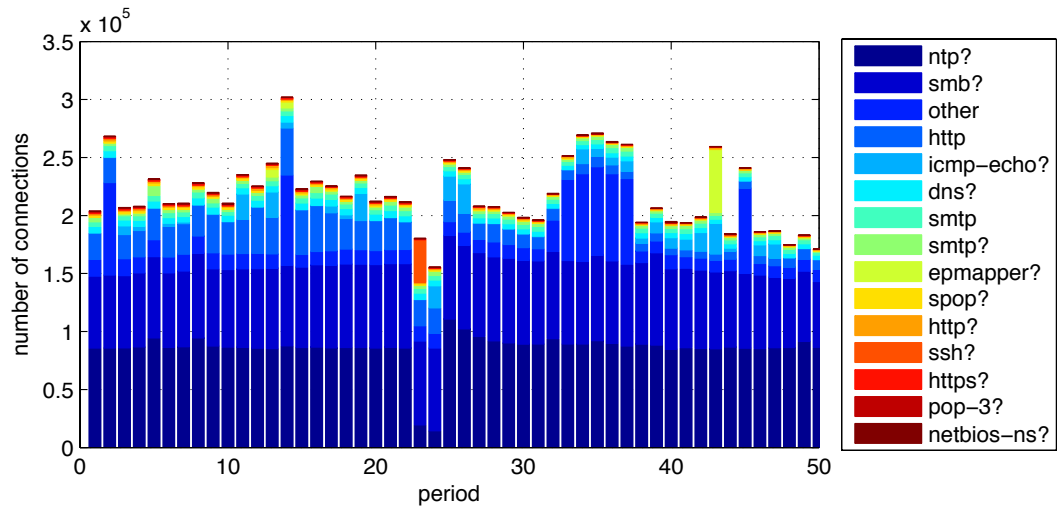

Figure 3.8 – Net-3. Outbound Connections

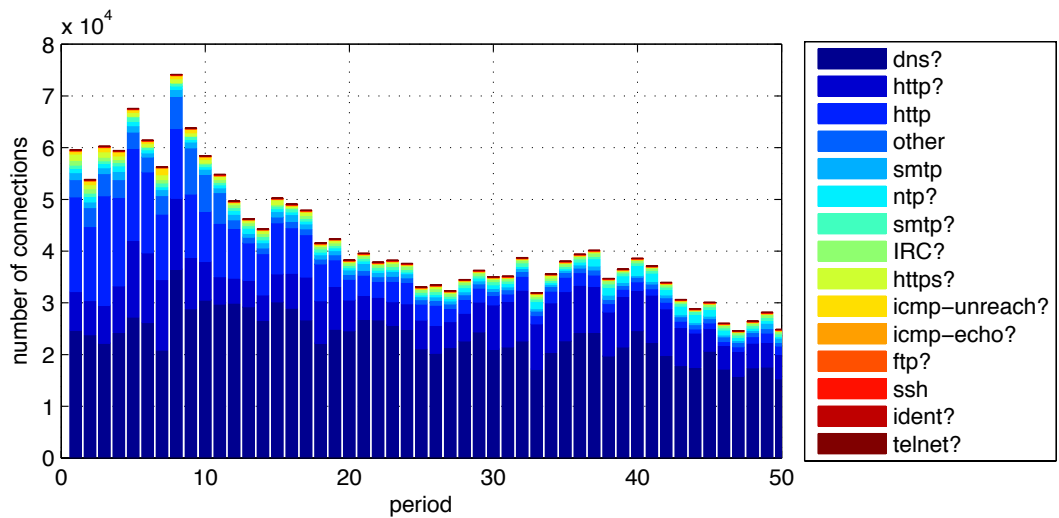

Table 3.4 – Amount of Traffic Transferred by Main Applications

| | Net-1 | | Net-2 | | Net-3 | |
|---|---|---|---|---|---|---|
| | In | Out | In | Out | In | Out |
| Amount of Traffic Transfer [GB] | 374 | 82 | 172 | 159 | 910 | 548 |
| Main Applications [%] | | | | | | |
| HTTP | 32.3 | 46.7 | 67.3 | 67.2 | < 1 | < 1 |
| OTHER | 8 | 36.3 | 14 | 12 | < 1 | < 1 |
| SMTP | 52.2 | 6 | 8 | 10.8 | 99.1 | 93.6 |
| FTP | 2.5 | 6 | 6.4 | 4 | < 1 | < 1 |
| HTTPS | < 1 | < 1 | 2.6 | 2.1 | < 1 | < 1 |
| SSH | < 1 | < 1 | < 1 | < 1 | < 1 | 2.3 |

## 3.4.2   Amount of Data Exchange

In order to understand the application behavior, we look at the bytes transferred by local and remote hosts in inbound and outbound connections registered in connection summary log. Our metrics discriminate between the traffic transmitted in normal established and terminated connections, flagged as "SF", and the rest of connections. In TCP connections, transmitted bytes are calculated using the header field "Sequence Number". It should be noticed that problematic connections might have erroneous values in transmitted bytes. Besides, in the cases that Bro did not see the end of the connection, i.e. a connection flagged with "S1", there are not byte counts inside the connection summaries. Even these limitations, we use these measurements as a first attempt to understand the nature of the traffic in each network. We summarize the results from incoming and outgoing traffic (we considerate in each category the traffic transferred in local and remote "SF-labeled" connections) in Table 3.4.

We expected to find more traffic transferred in HTTP outgoing connections; the difference is explained with the high traffic found in the Bro category "OTHER". Protocol violations and other anomalies affect the classification.  The transmitted bytes of Net-3 are highly biased, showing a 99 percent of SMTP traffic. By manually checking some SMTP connections in *Bro connection summaries*, we found 255 inbound connections where originator is sending between 1 and 2 Gb and the responder traffic is 0 bytes. A deeper investigation showed that the problematic connections involved three local IP addresses. Checking the *Bro weird.log* file for one of this IP disclose a wide variety of problematic TCP packets: SYN packets that contains data, TCP Christmas, RST storm, bad TCP checksum, bad TCP header length, SYN after RST, etc.

Table 3.5 – Amount of Traffic Transferred in Net-3 with correction applied

| | Net-3 | |
|---|---|---|
| | In | Out |
| Amount of Traffic Transfer [GB] | 8 | 38 |
| Main Applications [%] | | |
| HTTP | 78 | 47 |
| OTHER | 6 | 4 |
| SMTP | 1 | 10 |
| FTP | 2 | < 1 |
| HTTPS | 2 | 1 |
| SSH | 5 | 33 |

To correct the erroneous statistics we filter the three local IP addresses. New metrics for Net-3 are showed in Table 3.5. Filtering all the connections from one host is just a basic step and not the best solution. The lack of previous statistics of normal traffic transfer to use as ground truth makes this study the first attempt to characterize Todai traffic. More systematic study is needed to assure that statistics obtained using Bro records are correct. Anomalies in the traffic generate erroneous or missing values in the count bytes field of the connection summary. Possible values are "?", "0" or an excessive large number when the value is taken from the ACK field of a packet that announces "TCP Acked lost segment" (an acknowledge that Ethereal detects but cannot see the segment sent).

The amount of data exchange in a connection is used frequently in machine learning based research. The field "source bytes" and "destination bytes" are part of the basic features of the KDD benchmark dataset and the most discriminating feature for the majority of the classes (normal and attack connections) [40]. This situation raises doubts about studies using transmitted bytes; results among researchers are difficult to compare or validate without an adequate treatment in pre-processing phase and a documentation of the process. In the case of storing huge amount of data for posterior analysis, statistics and normalizing procedures would also be affected. This problem may not be restricted to Bro, but also appear in other systems that calculate the amount of data exchange.

## 3.5   Bogus connections

Connections that cannot be classified in any well-known service are labeled "other". In Net-1, this category represents 40 percent of the total incoming connections. A non-negligible 37 percent of this 40 percent are connection attempts with no reply. Bursty traffic on period 5, 17, 36 and 40 is generated by these bogus connections. While the number of normal established and terminated connections ("SF" label), which represent 42 percent of the incoming connections, steadily decrease as time passes, these bogus connections remains constantly with peaks of activity. A consequence of these partial connections is a dramatically decrease in well-established connections classified as "other" displayed in Figure 3.9. This phenomenon is explained by SYN-Floods [41], one of the most common denial-of-service attacks. Looking closer at the way SYN-floods are conducted over period 5 showed that duration span is approximately 2 minutes and the attack is repeated several times.

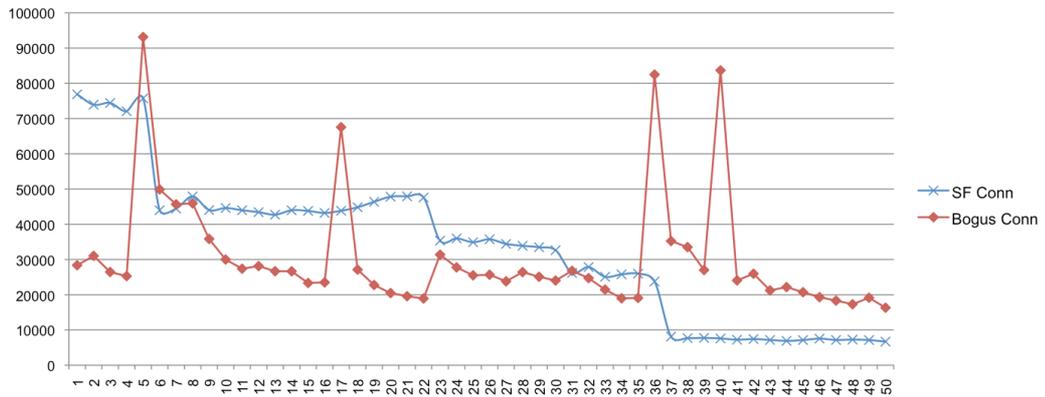

# Chapter 4

# Network Anomalies

## 4.1 Anomalies Overview

Internet traffic exhibits a wide diversity of behaviors. When sensor monitors analyze data, unusual events are often observed. Some of these events may indicate failures in network components, broken protocols, bugs in applications while others may indicate malicious actions to confuse the monitor and evade detection. Then, it is necessary to establish diagnostic criteria and define how to treat these "exceptional" events.

Bro IDS philosophy is to record all the events encounter by the analyzers because they can reflect erroneous assumption either by the application itself or by the users. Each event is associated with a customizable policy that dictates the action to take when the event engine generates it; see Table 4.1. While the general approach is to log all of them, in some cases errors occur in cascade, e.g. retransmission inconsistencies due to a TCP protocol bug. In those cases, the "weird_notice_once" policy filters the subsequent non-informative events.

For an initial deployment, the effort to understand unusual activities will provide the security practitioner an inside view of the network. However, from the point of view of usability the burden of analyzing thousands of events written in plain text files without knowing whether these records hold valuable information is a big concern when employing IDS [35]. It is common to read in the Bro mailing list that security practitioners ask methods to avoid recording this kind of anomalies. A solution based on machine learning methods is to reduce the number of alerts according two approaches

Table 4.1 – How to respond in the presence of unusual or "weird" events [30].

| Policy | Action |
|---|---|
| WEIRD_UNSPECIFIED | No action specified. |
| WEIRD_IGNORE | Ignore the event. |
| WEIRD_FILE | Record the event to weird file, if it has not been seen for these hosts before. (But see weird do not ignore repeats.) |
| WEIRD_NOTICE_ALWAYS | Record the event to weird file and generate a notice each time the event occurs. |
| WEIRD_NOTICE_ONCE | Record the event to weird file; generate a notice the first time the event occurs. |
| WEIRD_NOTICE_PER_CONN | Record the event to weird file; generate a notice the first time it occurs for a given connection. |
| WEIRD_NOTICE_PER_ORIG | Record the event to weird file; generate a notice the first time it occurs for a given originating host. |

[42]. Fusion is the combination of alerts representing independent detection of the same attack. Correlation is the clustering of alerts to provide a high level view of the intrusion attempts.

In order to apply machine learning wisely, a first stage in the research is the analysis of data. Even if the research project is aimed at understanding a specific aspect of security related topic, an overall examination is required to avoid biased results. The examination of anomalies in Todai trace showed that academic environments are replete with noisy sources that in many cases correspond to experimental systems. Measurements based on capture traffic generated on experimental systems may distort statistics.

It is hard to know how frequent are the anomalous events observed by Bro in other environments. Compare to the reference given in the Bro wiki page, the number of different unusual events in one night log from University of Tokyo surpasses the 42 distinct events seen in 10 month traces from the Lawrence Berkeley National Laboratory. In U-Tokyo environment, we found 62 types of anomalies; Appendix A includes plots for all of them. Each plot presents the number of records found in each of the 50 analyzed periods discriminated in traffic direction and IP address block.

Bro classifies anomalies according its origin: network behavior (when the event cannot be associated with a pair of hosts), flows (a pair of host can be identified); and connections (time, source IP, destination IP, source port and destination port is available).

## 4.2   Dissecting Anomalies

### 4.2.1   Unusual Flows

This type of anomalies is triggered when the event cannot be associated with any connection, only with a pair of host corresponding to a flow between source address and destination address.

All events deal with unusual IP fragments that in general are allowed by the standard but are unlikely to happen. TCP/IP suite supports the fragmentation of large packets that otherwise cannot be transmitted in a data link due to the link maximum transmission unit (MTU). To allow reassembly process, fragments include an identification field to indicate the original datagram (non fragmented packet), a fragmentation offset to indicate the position in the original datagram, and a flag to indicate the last fragment.

Exploits to IP fragmentation are generated to bypass security measures or for denial of service attacks (DoS attacks). Indicators of the attempts to evade the monitor are "fragment inconsistency", "excessively small fragment", or "fragment size inconsistency" events. Examples of the latter case are the use of fragmentation overlap, which is the basic form of teardrop DoS attacks, or large IP fragments to crash some IP implementations. However, there is also a possibility of innocuous fragmentation events, sometimes fragments may overlap old fragments that were not flushed from Bro cache.

In Todai trace, a low presence of fragmentation problems in each of the three networks is observed. Events observed are excessively_large_fragment, excessively_small_fragment, fragment_with_DF, fragment_inconsistency, fragment_size_inconsistency and fragment_overlap. When checking manually, we detected that some of the fragments that triggered inconsistencies and overlapping problems do not posses an appropriate value in the identification value and/or datagram offset. Reassembly of these fragments is not possible. The appearance of these events is associated with DoS attempts.

**DoS attack**

An exception occurs during periods 11-14 in Net-2. From 21:23:40 PM to 22:13:06 PM, the number of incoming connections attempts (with no reply seen) in each period increase from around 90,000 to 300,000. Rejected connections also increase from to 2,000 to 30,000. The number of normal established and terminated connections remains constant. A burst of probes, all coming from the same source IP address target ports: 80, 1080, 3124, 3127, 3128, 6588, 8080, 50050. Port 3127 is a favorite backdoor created by myDoom variants and more recently Agobot family. During the attack, an outstanding quantity of "fragment_inconsistency" events is observed in the *weird.log* file. In period12, only one remote IP address and one local IP address are involved in 1844 events represents 70% of the total highly unusual fragmentation inconsistencies observed.

Using Bro, this attack can be observed by the analysis of *conn.log* and *weird.log* files. The lack of hints to recognize this attack inside the *alarm.log* file points out the importance of analyzing anomalous events that do not necessary trigger alarms.

DoS attack is confirmed by the Frag3 preprocessor (ssp_frag3) of Snort. Alarms triggered on period 12 are: Fragmentation overlap (# 2168), Teardrop attack (#5) and Tiny fragment (#1). Teadrop attack exploits IP fragmentation using crafted fragments with false information on the offset field to cause the system to be unstable during packet reassembly. During the attack the number of normal established connections does not decrease, indicating that recent systems are no longer vulnerable to this attack. Results are limited by the offline analysis; in online analysis a flood of fragments may saturate IDS resources to make a successful attack.

**Overlapping Fragments**

It should be noticed that in period 12, Bro raised a total of 2615 fragmentation inconsistencies; the gap between Bro fragmentation_inconsistency events number and Snort ssp_frag3 alarms number is probably due to configurations on both systems.

IDSs do not necessary observed the same traffic that endpoints see. Attackers may manipulate its own traffic to exploit flaws existing in several stages of the packet evaluation and reassembly processes. Description of denial of service, insertion attacks (when monitor analyze packets that will never reached the destination) and evasion attacks (when monitor fail to observe some packets) are given in [43]. This problem affects IP fragmentation as well as TCP segmentation reassembly processes. When retransmission and data overlapping occur, some systems favor old data while others favor new data.

Shankar and Paxson draw attention to five methods for fragmentation reassembly performed by different operating system [44]. Snort engineers focus on TCP fragmentation conflicts and reveal more possible policies in [45] making them available for users to customize preprocessors in the configuration file. Snort incorporates the use of a Host Attribute Table to specify a policy per each host of the local network and a general policy for hosts not listed on the table. Such table is design for improving the output of Snort frag3 and stream5 preprocessors, which are target-based system; output depends on the directives passed to the system. In a large network environment, where NAT servers hide inner hosts and DHCP protocol frequently reassigns IP addresses, the Host Attribute Table is an unmanageable solution and other strategies are needed.

For this research, the Snort default reassembly policy is set to Windows meaning that Snort will reassembly packet as its destination is a Windows host. Windows favor old segments, except when the subsequent segment begins before the original segment.

Bro does not support policy configuration. In the case of online analysis an active mapping for fingerprinting host operating systems was available but it is no longer supported. For the case of TCP segmentation overlapping, only the segments identified as malicious trigger rexmit_inconsistency events in the *notice.log* file. These events are associated with connections and are not part of this analysis.

### 4.2.2   Unusual Connections

This category corresponds to anomalous events that can be identified by the tuple <time when the event is seen, source IP, source port, destination IP, destination port>. Included events are handled by Bro conn_weird, conn_weird_add1 and other events generated by selected analyzers. Table 4.2 list all the unusual events observed in Todai trace. Some of them refer to problematic packets, others to general characteristics of the connection and other to application specifics.

It is observed that the likelihood of these events depends on the direction of the connection, the observed time and the network. Besides, some anomalies are presented in bursts, others appear steadily during the whole period and others decrease frequency as time progresses. Sometimes, the appearance of one or two anomalies is a clue to predict the arrival of others events. Previous work showed correlations among anomalies [46].

Table 4.2 – List of unusual events associated with connections

| | |
|---|---|
| above_hole_data_without_any_acks | simultaneous_open |
| data_before_established | partial_ftp_request |
| excessive_data_without_further_acks | malformed_ssh_identification |
| FIN_advanced_last_seq | partial_RPC_request |
| FIN_storm | unpaired_RPC_response |
| inappropriate_FIN | inflate_failed |
| possible_split_routing | NUL_in_line |
| connection_originator_SYN_ack | HTTP_chunked_transfer_for_multipart_message |
| SYN_seq_jump | HTTP_version_mismatch |
| SYN_after_reset | base64_illegal_encoding |
| bad_SYN_ack | illegal_%_at_end_of_URI |
| SYN_inside_connection | unescaped_%_in_URI |
| SYN_after_close | unescaped_special_URI_char |
| SYN_after_partial | double_%_in_URI |
| SYN_with_data | unmatched_HTTP_reply |
| repeated_SYN_with_ack | unexpected_multiple_HTTP_requests |
| RST_storm | line_terminated_with_single_CR |
| TCP_christmas | HTTP_bad_chunk_size |
| baroque_SYN | can't be parse |
| corrupt_tcp_options | multiple_login_prompts |
| bad_TCP_header_len | no_login_prompt |
| window_recision | irc_line_too_short |
| bad_UDP_checksum | irc_invalid_command |
| bad_ICMP_checksum | irc_invalid_topic_reply |
| bad_TCP_checksum | irc_invalid_line |
| premature_connection_reuse | irc_invalid_reply_number |
| data_after_reset | irc_invalid_who_line |
| active_connection_reuse | irc_line_size_exceeded |

There are various reasons that can explain why anomalies occurred. To simplify the complex analysis of more than ten terabytes, a principle from the economist Vilfredo Pareto is imported to the information security field. Pareto Principle, "80-20 rule", states that roughly 80 percent of the effects come from 20 percent of the causes. In computer science, this principle applies to some aspects of software engineering and recently a study to predict attack-prone components in a system was published [47].

Activity in the academic network includes controlled experiments that can be source of hundreds, or sometimes thousands of anomalies over a short period of time. In some cases, the experimental hosts transport benign anomalous traffic that may not raise attack alarms but posses unusual traffic characteristics. In other cases, the inherent characteristics of the research project, e.g. honeynets or anonymous proxy servers, make the system vulnerable and up to 100% of traffic could be considered malicious. These noisy systems explain a large number of anomalies observed in Todai traffic. While the 80-20 rule is not strictly followed, a large percentage of anomalies can be associated with a small number of hosts.

The inclusion of disruptive sources in datasets aimed at anomaly detection research should be studied carefully since they can undermine statistics and affect the visibility of network traffic diversity.

When observing anomalous events over a long time period, in our experiments several hours, the recognition of noisy sources is easier than in online analysis. Then, a good practice for conducting online analysis is a prior offline assessment of network characteristics.

In this section, examples to illustrate anomalous behaviors in U-Tokyo Network are presented for packets, TCP connections and applications. IP addresses that generate an annoying quantity of events are identified as sources of noise. The anomalies selected represent more than one third of the total anomalous events triggered by Bro. Snort alerts are used for better understanding when is possible. Each anomaly is summarized in a table, indicating the amount of events, a ranking[2] number (RK) indicating how frequent the event is, and the amount of events that can be reduce to few sources of noise.

**Packet Anomalies**

**bad_ICMP_checksum**

| Description | *This event indicates that the checksum field in ICMP packet was invalid.* | | | | |
|---|---|---|---|---|---|
| Observed # | **7,259 (RK 14)** | | | | |
| Origin | **65%** | **targets** | **53** | **net-1** | IP addresses |
| | **80%** | of this | **65%** | Involve | **1 local IP** |
| | **30%** | **targets** | **53** | **net-2** | IP addresses |
| | **76%** | of this | **30%** | Involve | **2 local IP** |

Table 4.3 – Statistical summary for bad_ICMP_checksum

---

[2] RK 1 means most frequent event

Top rank affected hosts are analyzed. Less than 5 hosts in each local net are the major destination of the ICMP packets. The two most affected IP address of net-2 correspond to two machines that are used for Planetlab [48], a testbed system open to public. There is no information about the uses of hosts from net-1.

Encrypted connections and connections classified as "other" over tcp and udp protocols are found. In both nets the ICMP packets are answers to outbound connections attempts over UDP to unreachable remote ports. Bit Torrent may have generated this pattern in traffic.

Data transmitted inside the verified ICMP packets correspond to the encapsulated UDP packet that generated the ICMP message. No particular problems are observed.

Snort Preprocessor Stream5 does not contribute with more anomalies.

It should be notice that the possibility of carrying data in ICMP packets may be exploit by attackers to carry data surreptitiously. Especially the ICMP echo-request/echo-reply are use to create convert channels. In some situations, the simplicity of ICMP packets is seen as an advantage. The lack of port numbers may be used to bypass firewalls or other security measures. Botnets use ICMP tunneling to establish communications with the compromised machines. Distributed Denial of Service attacks (DDoS) targeting Amazon, Yahoo and other large companies have been organized through ICMP tunnels [49].

IDS systems are challenged to differentiate between legal and illegal ICMP packets. Recognized encrypted and non-encrypted payload is also an open problem.

**bad_TCP_checksum**

| Description | *Invalid TCP checksums might be used in subtle attacks where an attacker is aware of the presence of a monitor between them and the victim machine and tries to convey their activity undetected* | | | | |
|---|---|---|---|---|---|
| Observed # | **151,413 (RK 3)** | | | | |
| Origin | **67%** | **initiated by** | **203** | **net-2** | IP addresses |
| | **74%** | of this | **67%** | involve | **10 local IP** |
| | **97%** | of this | **67%** | **target** | **port 20480** |
| | **30%** | **initiated by** | **135** | **net-1** | IP addresses |
| | **80%** | of this | **30%** | involve | **7 local IP** |
| | **95%** | of this | **30%** | **target** | **port 20480** |

Table 4.4 – Statistical summary for bad_TCP_checksum

Top rank affected hosts are analyzed. Unclassified connections attempts to port 20480 are seen over the 50 periods. IP destination address belongs to Google, Akamai, NTT, TrendMicro, Yahoo, Microsoft among other large well-known companies and host are mainly located in U.S.A., Japan and China. Port 20480 is a register port for the emwave Message Service but connections observed in Todai traffic do not seem carry this protocol. Bro identifies this traffic with "OTH" status, which means "no SYN seen, only midstream traffic".

The examination of some packets using Wireshark brought out that flags "RST" and "ACK" are set and checksum is incorrect.

Possible reasons for initiating connections towards port 20480 are:
- Multiplayer Game Services
- Malformed packet towards port 80. Computers interpret numbers differently according the endianness of the processor architecture. Endiannes indicates the order to read individual subunits within a longer data word. Little endian machines will read port 80 as "50" in hex and "01010000" in binary. It also reads port 20480 as "5000" in hex and "0101000000000000" in binary.

Bugs in multiplatform applications may mislead port 80 to port 20480. We have seen some services running on DirectServer/8.0.7.i6 iPSX (Solaris; Ultra/Sparc), which is a big endian architecture. Some Sparc processors can access data in both endian orders, the option is at the application instruction level or at the memory page level.

**TCP Connection Anomalies**

**above_hole_data_without_any_acks**

| Description | *Could mean packet drop; could also be a faulty TCP implementation. This anomaly is worth to evaluate in online analysis, to detect packet dropping.* | | | | |
|---|---|---|---|---|---|
| Observed # | **156,227 (RK 2)** | | | | |
| Origin | **74%** | **targets** | **232** | **net-1** | IP addresses |
| | **68%** | of this | **74%** | involve | **2 local IP** |
| | **99%** | of this | **74%** | **target** | **1 subnet C** |

Table 4.5 – Statistical summary for above_hole_data_without_any_acks

The plot in Appendix A shows a high number of events at the beginning of the capture. It indicates that BRO was not able to see the ACK packets. When data exceeds the "max_above_hole_without_any_acks" threshold, an event is triggered and buffer is released. The increasing trend around periods 33-40 is due to a longer interval of time between analyzed periods. The gap between periods 32-33 is nearly 15 minutes and between periods 38-39 is about 19 minutes.

Top rank affected hosts are analyzed. The amount of incoming events for net-1 is significantly high. 73% of total events (99% of the incoming events to net-1) are mostly related to connections attempts from a laboratory in the U.S.A. to one local subnet class C, particularly to one host (68% of activity). That host is associated with a research project. However, connections status are mostly labeled as irregular: S0, SH and SHR.

Bro triggered 246 alarms of port scanning activity from the same laboratory.

**Application Anomalies**

**HTTP_version_mismatch**

| Description | *A persistent HTTP connection sent a different version number for a subsequent item than it did initially.* | | | | |
|---|---|---|---|---|---|
| Observed # | **25,030 (RK 12)** | | | | |
| Origin | **45%** | **initiated by** | **355** | **net-2** | IP address |
| | **72%** | of this | **45%** | Involve | **2 local IP** |
| | **96%** | of this | **45%** | Targets | **2 port** |

Table 4.6 – Statistical summary for HTTP_version_mismatch

72% of the total events are generated by two host from Planetlab that run an experimental http proxy service. In a 10 min of traffic capture, we found that 60% corresponds to "Bad Request". The activity of this server also generates other anomalies:

63% unexpected_multiple_HTTP_requests, 46% line_terminated_with_single_CR,

21% bad_ICMP_checksum, 19% unescaped_special_URI_char,

8% unmatched_HTTP_reply

## 4.3 Final Observations

The previous sections have served to explore in detail the most characteristic feature of the anomalies found using Bro. This research would not be complete without considering the events detected by Snort.

We applied Snort in 20% of the data analyzed with Bro. The outcome included an outstanding number of events related with fragmentation and overlapping (more than 90% of the events trigger are fragmentation). These anomalies combined with events of bad checksum in TCP packets (observed with Bro) can be connected with cases of evasion. However, limitations on the tools or in our analysis may generate a large amount of false alarms. Bro also identified an important amount of this type of events, but the fragmentation problem and the other events are more evenly distributed. A possible explanation could be the way the two systems process the data and the policies for fragmented packet reassembly. We think a study focused only on the fragmented packets would illuminate on this problem. Machine learning could help, e.g., packet header values can be used for clustering fragmented packets.

If these events are associated with evasion techniques used by botnets, a look at the output of SMTP and DNS Bro analyzers will also help in understanding this matter.

# Chapter 5

# Conclusion and Future Work

In this thesis, we presented the analysis of anomalies in the Internet traffic observed at the U-Tokyo campus network gateway. This chapter summarizes the outcome of our study and proposes ideas for future research.

## 5.1    Summary

IDS based on machine learning algorithms have low acceptance in the operational real-world network environment [2]. One of the causes is the use of data that failed to represent the actual behavior of Internet for training and validation of new models. In particular, a considerable part of the bibliography found on anomalous research is based on the outdated KDD dataset benchmark, which was generated in a simulated network environment. Our thesis is that the most important part of pre-processing is auditing (1) real, (2) up-to-date and (3) large amount of data before selecting reliable features for IDS. When using an already prepared dataset, researchers do not have the opportunity to face the limitations that appear when real data is analyzed. Besides, in the case of KDD data, studies are limited to 41 features that describe the traffic generated by user connections with the assumption that they can describe real, actual Internet traffic. New Internet traffic characteristics are not considered.

For this thesis research, we analyze 6.5 TB tcpdump binary data that represents 12 hours traffic of a large academic network environment. Academic network traffic has unique characteristics. To the best of our knowledge, this is the first attempt to construct a baseline of network traffic behavior that highlights a series of problems encountered in the analysis of a large amount of captured packets.

Our work made the following main contributions:

- We conducted an analysis that covers a wide aspect of anomalous behavior. This thesis presents diverse pathologies found at different levels of the network stack covering a wide variety of cases ranging from fragmentation to bad packet checksums to TCP anomalies and to application anomalies that represent 40% of the total anomalous events triggered by Bro IDS. Using the KDD dataset as a reference, we identified 5 behaviors that are missing from KDD dataset and could be defined as new features or can be used to rule out data that may distort statistics and consequently reduce the detection power of machine learning algorithms. These characteristics are: bugs in applications, detection of experimental systems (typical presence in academic networks), server found running on non-standard ports, protocol violations and indicators that may alert about evasion techniques.

- We detected DoS attacks. Two different types of DoS attacks are documented. The first case is generated by SYN flooding and the technique demonstrates to be effective in that scenario. The second case is generated by packets fragmentation. The latter may become a threat to exhaust IDS resources.

- We documented problems in research that may lead to wrong results due to misinterpretations of data or misconfigurations in software (IDS configuration "by default"). Participants with diverse backgrounds do research on network anomalies. Thus, our approach is trying to understand better the nature of common anomalies found in current operational networks to benefit researchers coming from network, information security and machine learning areas.

- We assessed the quality of data by analyzing the potential and the real problems in the capture process. The evaluation also beneficiates the other experiments run in our department. The capture machine manufacturer has been informed about our findings. Bugs are subject to deep investigation.

- We described the application breakdown in terms of connections and the amount of traffic transfers. Both approaches show different aspects of the network behavior and contribute to a better understanding of the service and applications found.

## 5.2   Future Work

During this research, we have identified several opportunities for extending our work:

**Fine-grained analysis**

- A coarse-grained analysis is necessary to detect a wide variety of anomalous events and generates a baseline to show the interesting and curious patterns that describe U-Tokyo network behavior. However, a fine-grained approach is needed to focus on threats such as botnets and distributed attacks. Currently, data is also used for research on spamming and video streaming.

- We analyze 10% of the anomalies triggered by BRO that represents 34% of the total events. In Appendix A, plots for the totality of triggered events show how frequent anomalies are found. Exhaustive analysis of the causes is also interesting.

**Comparisons**

- Lax security measures and experiments may cause academic networks to show to some extent a different behavior than company networks. We need to have a baseline for commercial networks to measure how we depart from security problems observed on business environments by using data capture in the academic network.

- We constructed a baseline based on U-Tokyo network behavior over the night of a weekday in July 2009. Network behaves differently according the hour of the day, the day of the week and the month of the year. To find invariants in the network behavior, it is necessary to examine the same network using up-to-date data. For understanding state-of-the-art attack techniques and the evolution of evasion methods to bypass security controls, we also need to compare and contrast the outcome of this thesis with the analysis of a new capture.

- Preprocessors in Snort IDS identify some network anomalies. To conduct our research, we run Snort to improve our understanding on some particular problems covering only 20 percent of the captured data.

**Machine Learning**

- To achieve our goal, there is a huge amount of manual processes that should not be underestimated. Each environment has the unique characteristics and researchers need to learn them before using data for metrics. Then we need a framework to facilitate these operations. Simply clustering sources causing noise for each anomaly may ease our efforts. Features for clustering may be the IP destination, the source port and the time gap between events among other variables ranging from one anomaly to another.

- An open topic for research on machine learning is finding connections between traffic analysis and anomalies, for example, determining the network conditions that precede the occurrence of anomalies.

- To detect evasion techniques, we need to study in more detail header fields of the packets involved in these anomalies. For example, inconsistent values of TTL may indicate attacks.

**Appendix**

**A   Frequency of anomalous events**

**1. unmatched_HTTP_reply**

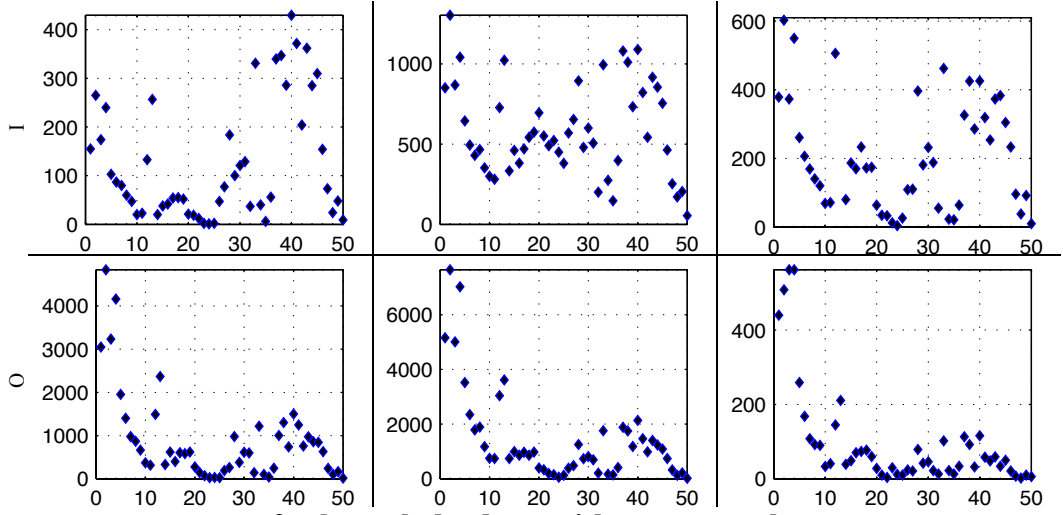

**2. above_hole_data_without_any_acks**

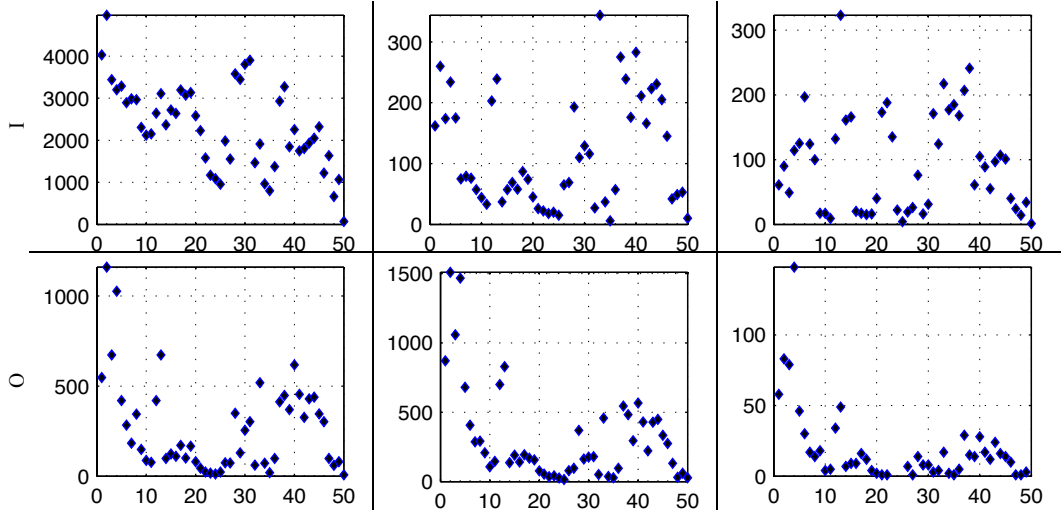

**3. bad_TCP_checksum**

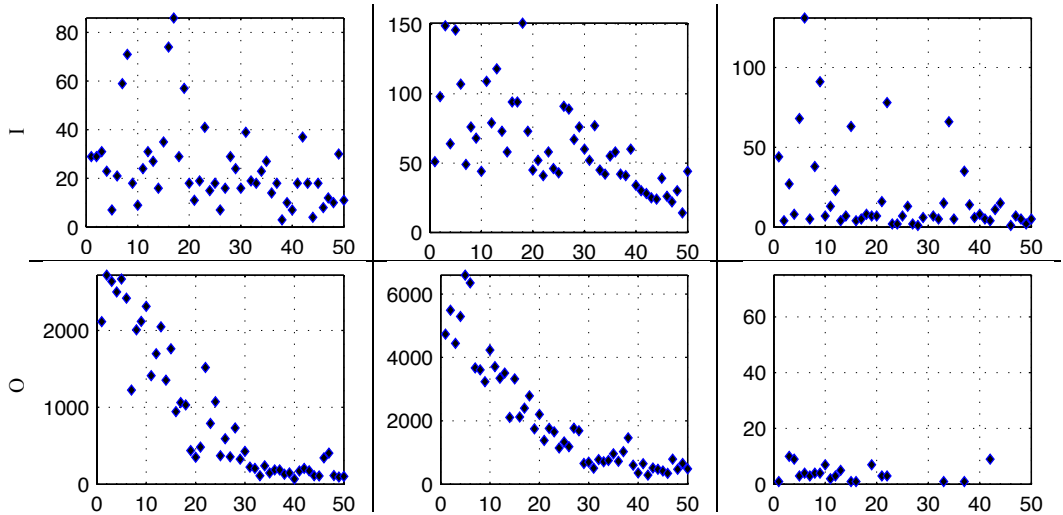

x-axis: 50 periods, y-axis: Number of anomalies found on that period. Left: NET-1, center: NET-2, right: NET-3

## 4. unescaped_special_URI_char

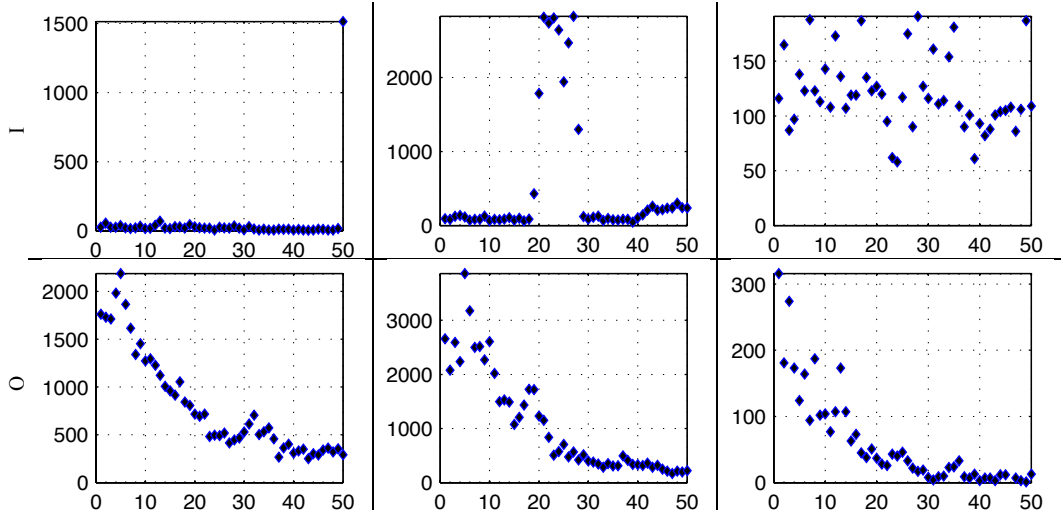

## 5. can't parse

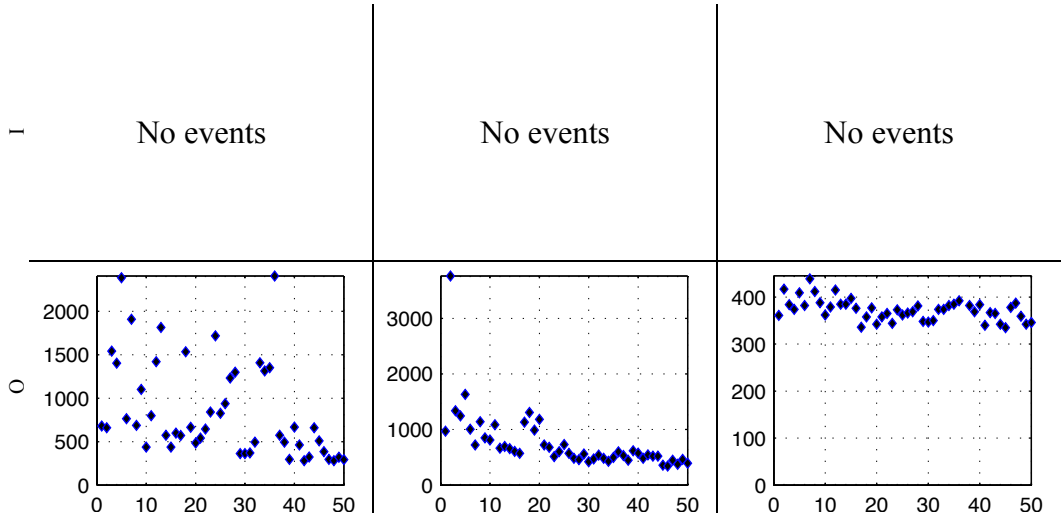

## 6. line_terminated_with_single_CR

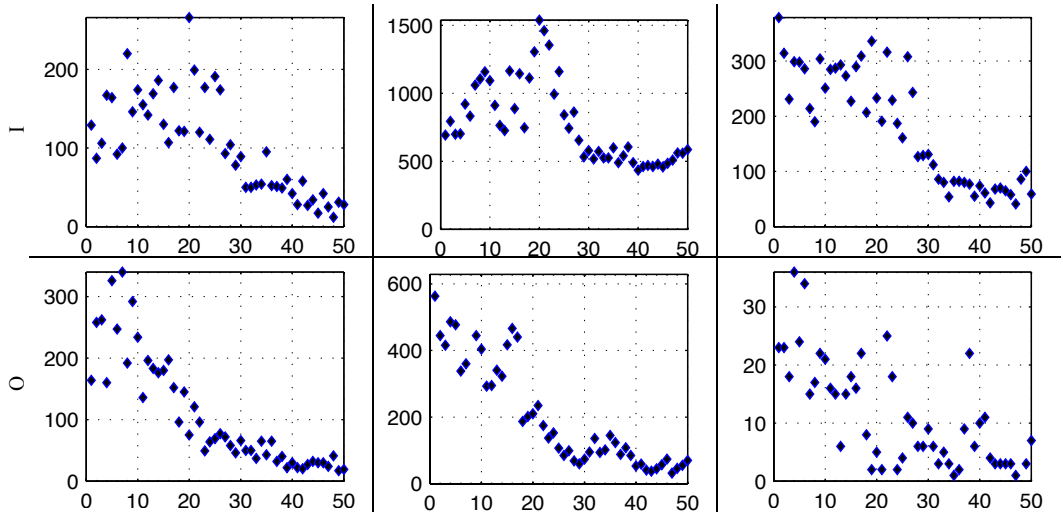

x-axis: 50 periods, y-axis: Number of anomalies found on that period. Left: NET-1, center: NET-2, right: NET-3

## 7. unescaped_%_in_URI

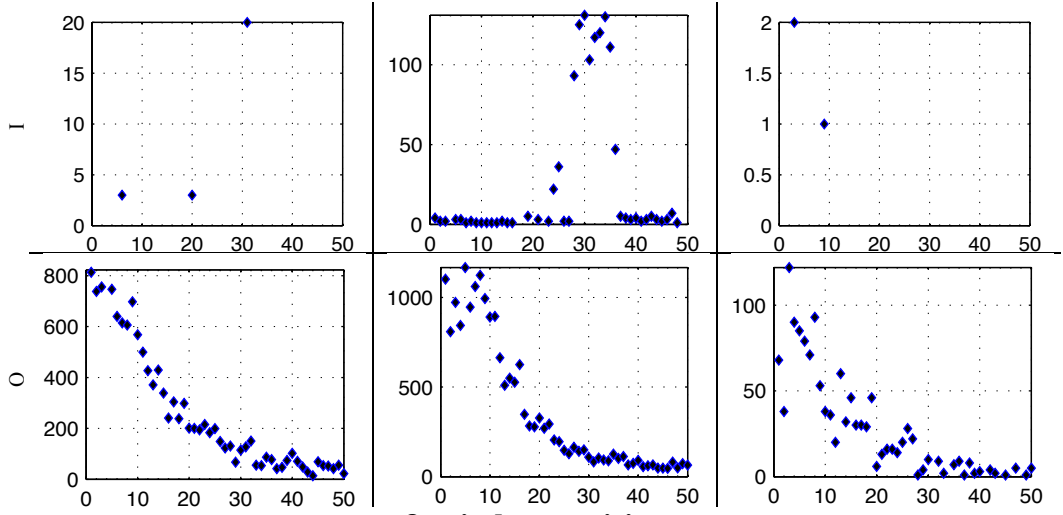

## 8. window_recision

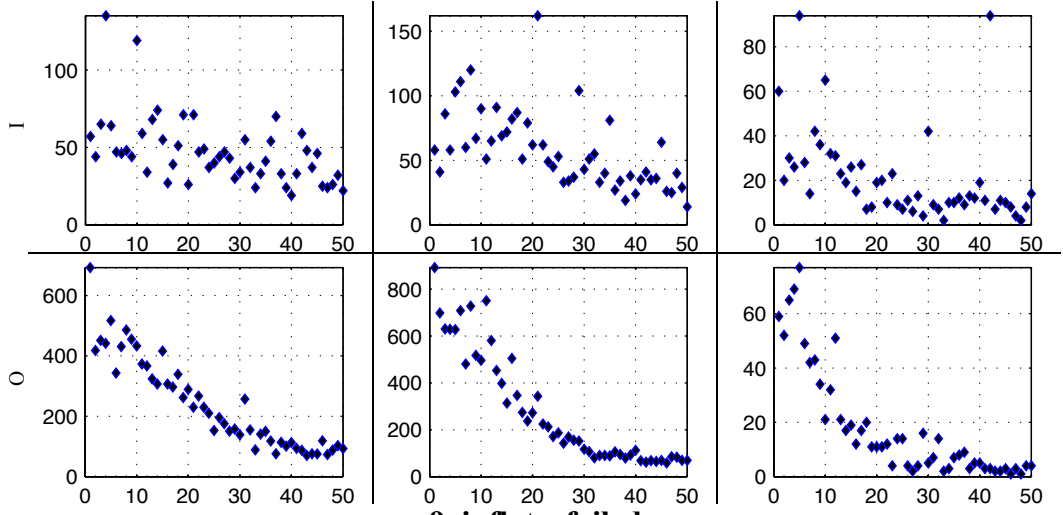

## 9. inflate_failed

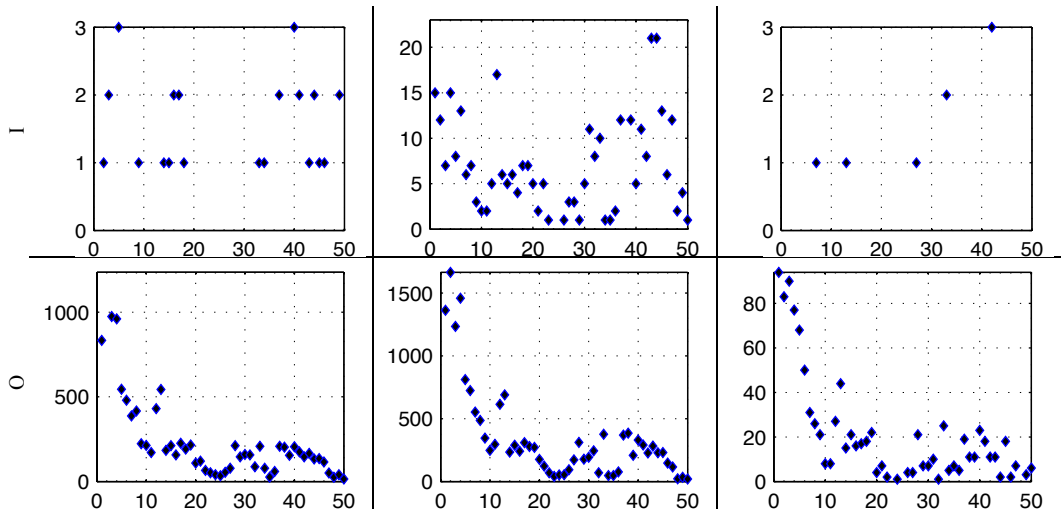

x-axis: 50 periods, y-axis: Number of anomalies found on that period. Left: NET-1, center: NET-2, right: NET-3

## 10. data_before_established

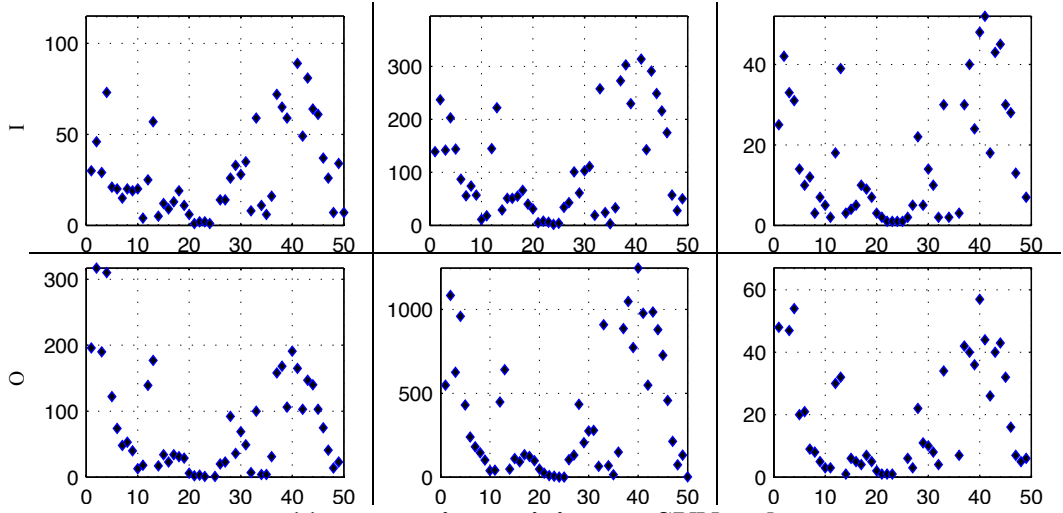

## 11. connection_originator_SYN_ack

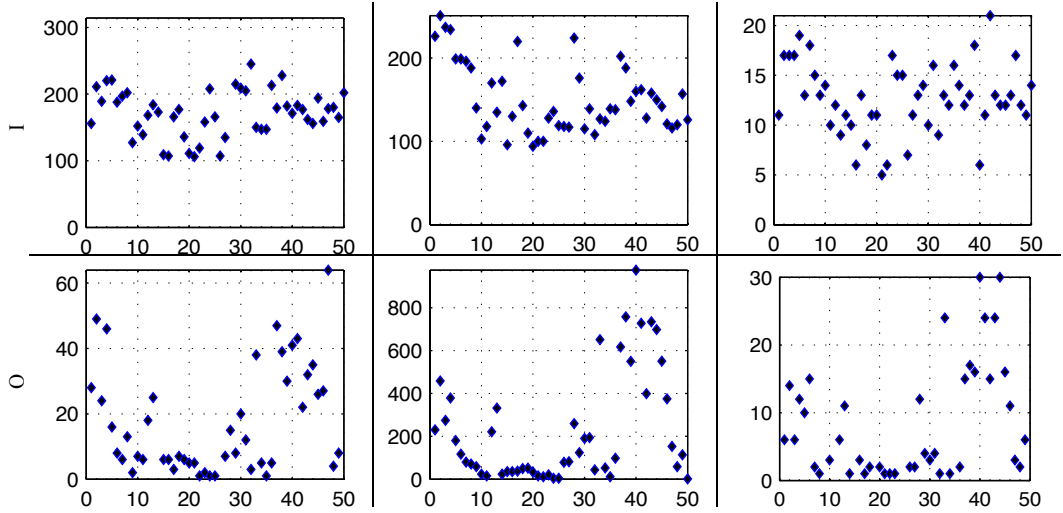

## 12. HTTP_version_mismatch

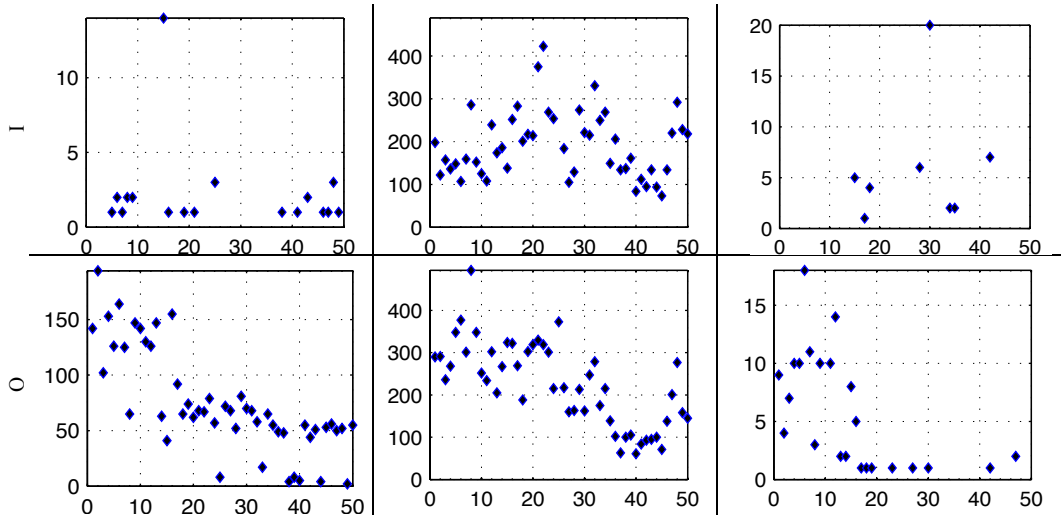

x-axis: 50 periods, y-axis: Number of anomalies found on that period. Left: NET-1, center: NET-2, right: NET-3

## 13. data_after_reset

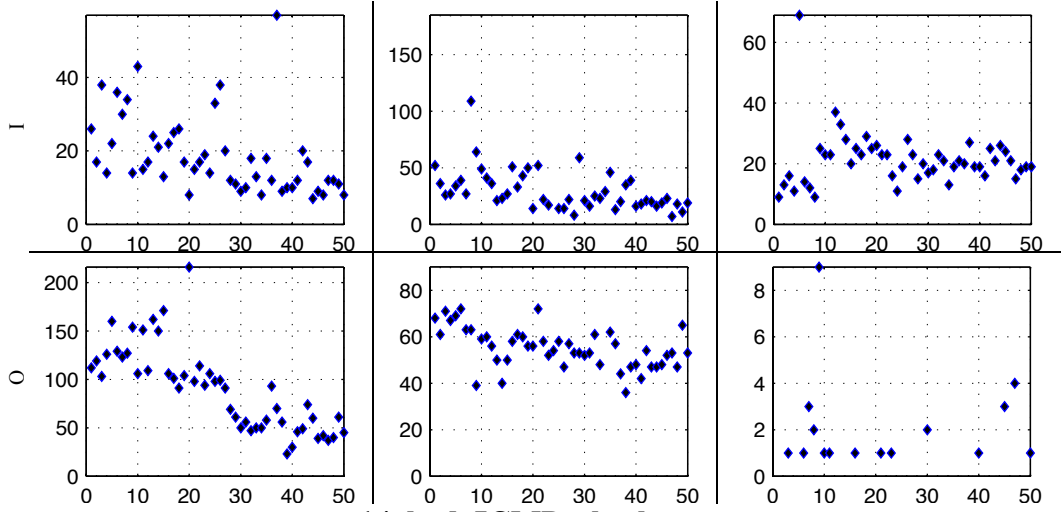

## 14. bad_ICMP_checksum

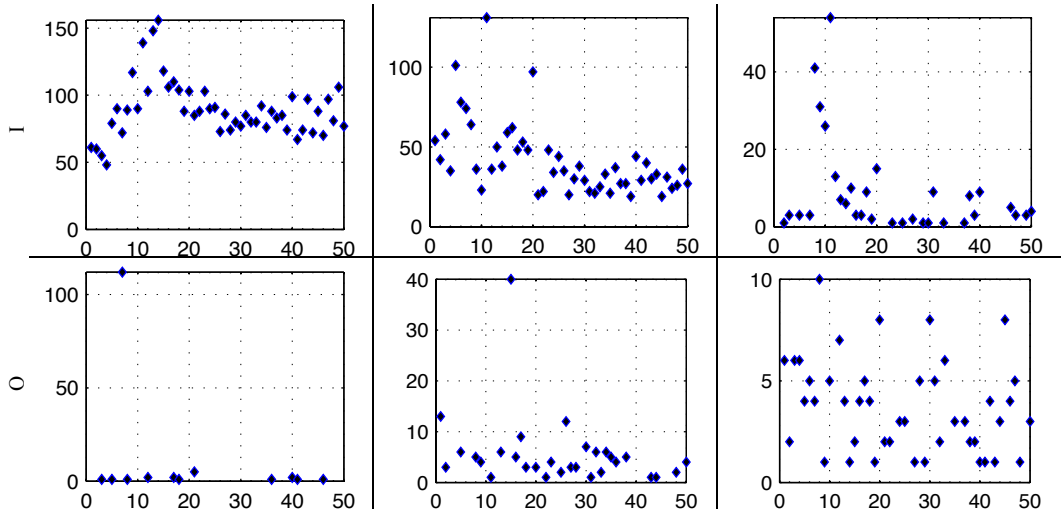

## 15. bad_UDP_checksum

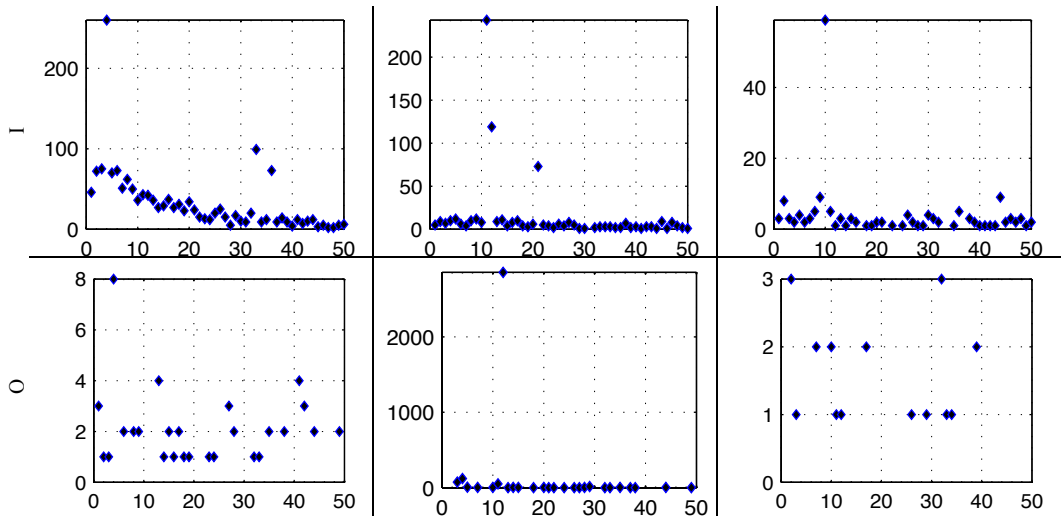

x-axis: 50 periods, y-axis: Number of anomalies found on that period. Left: NET-1, center: NET-2, right: NET-3

## 16. SYN_with_data

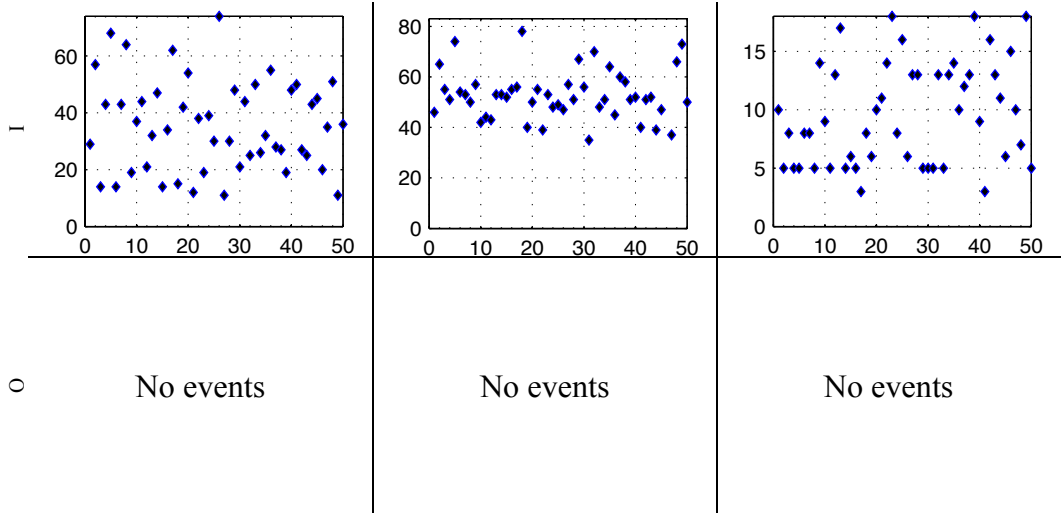

## 17. unexpected_multiple_HTTP_requests

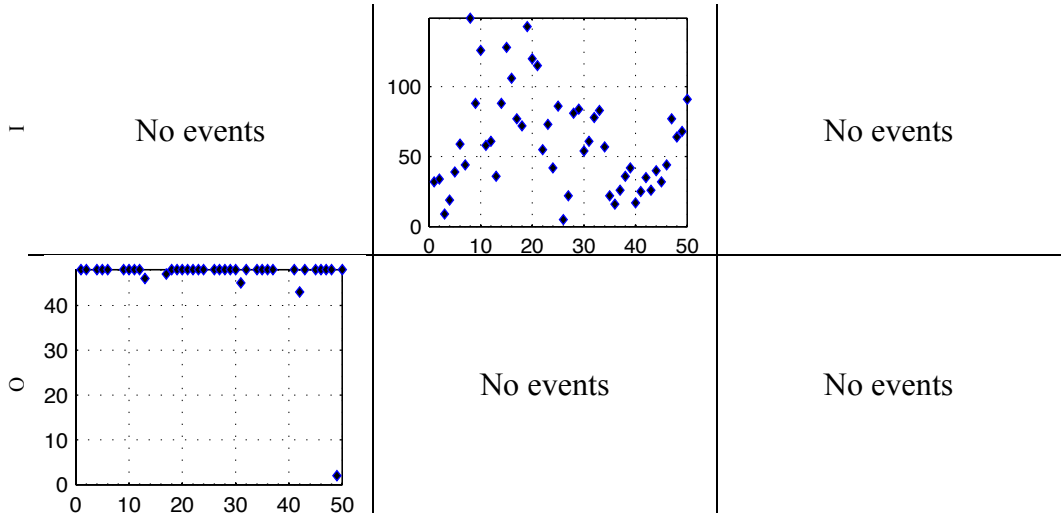

## 18. possible_split_routing

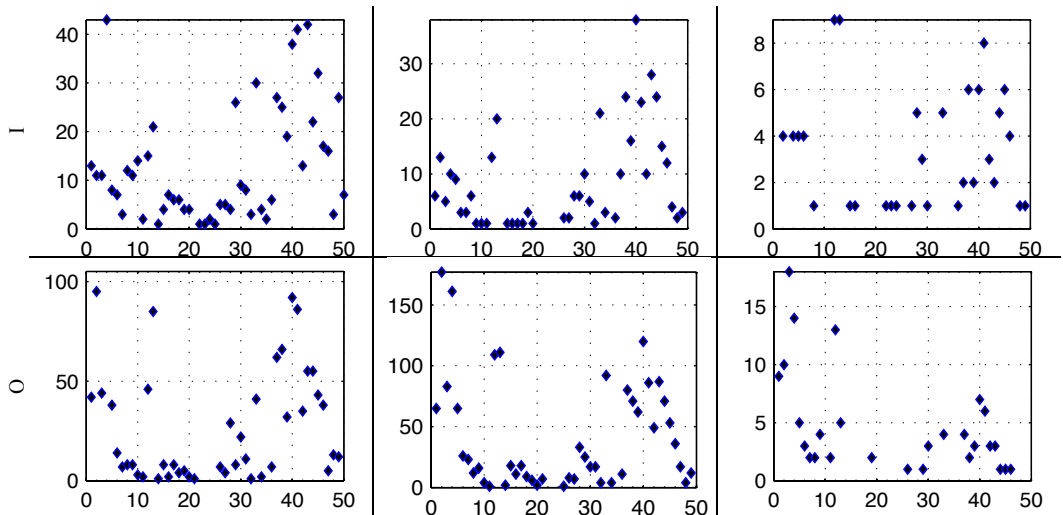

x-axis: 50 periods, y-axis: Number of anomalies found on that period. Left: NET-1, center: NET-2, right: NET-3

## 19. SYN_seq_jump

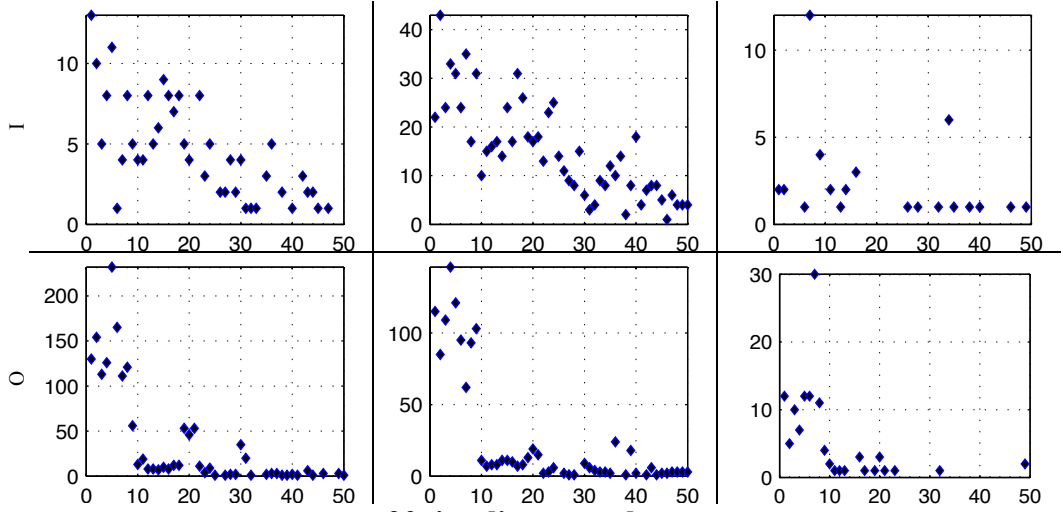

## 20. irc_line_too_short

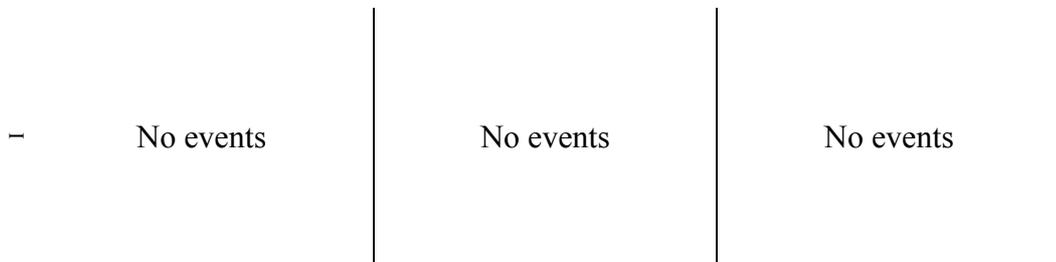

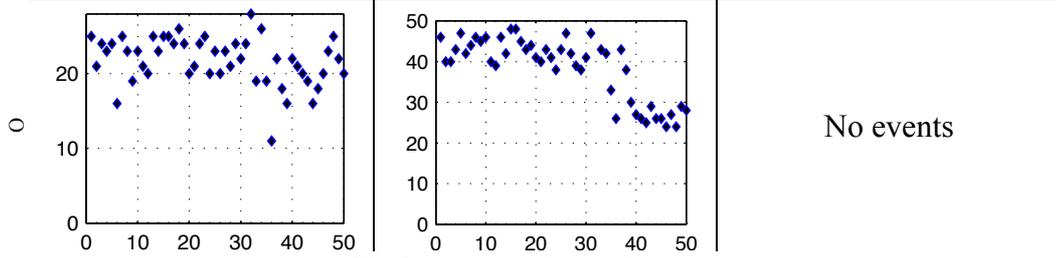

## 21. fragment_inconsistency

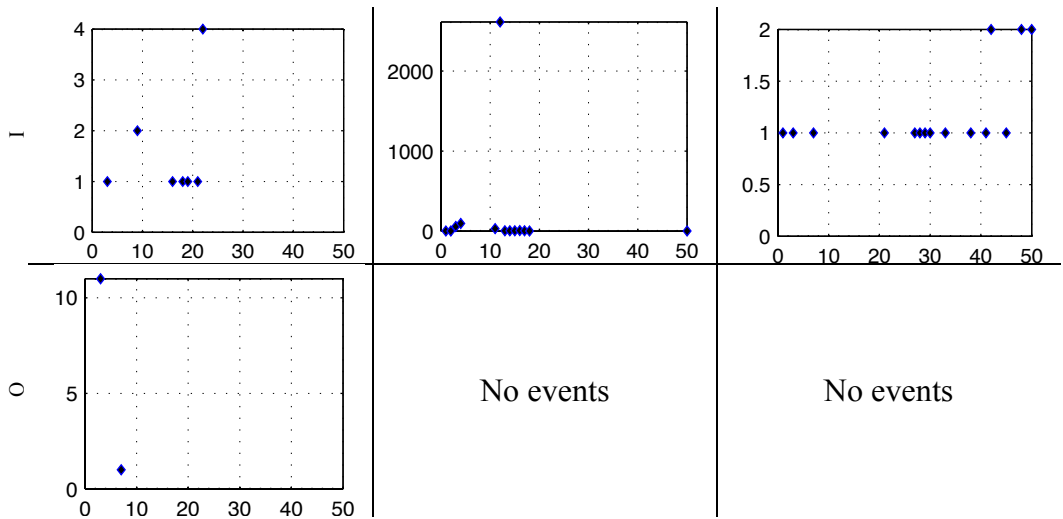

x-axis: 50 periods, y-axis: Number of anomalies found on that period. Left: NET-1, center: NET-2, right: NET-3

## 22. base64_illegal_encoding

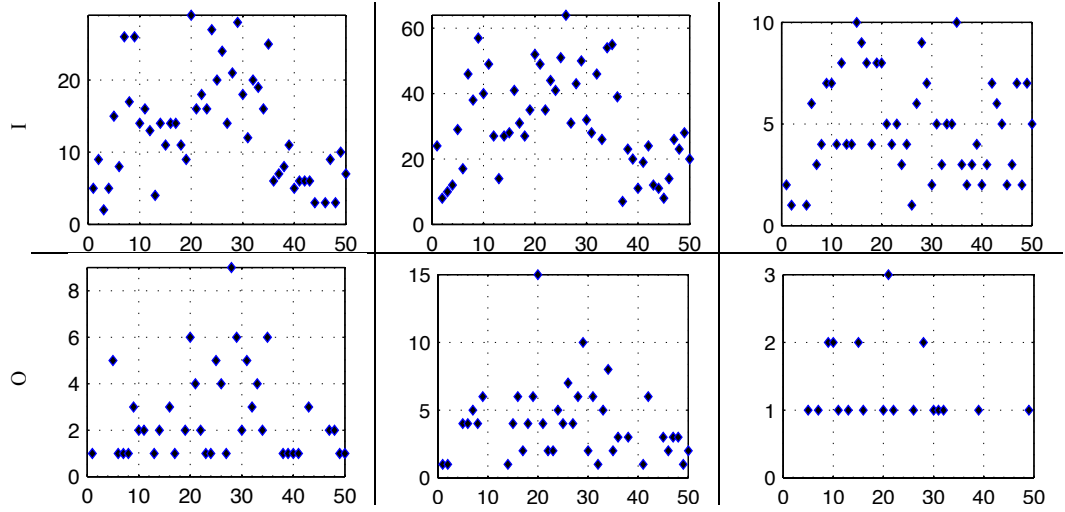

## 23. NUL_in_line

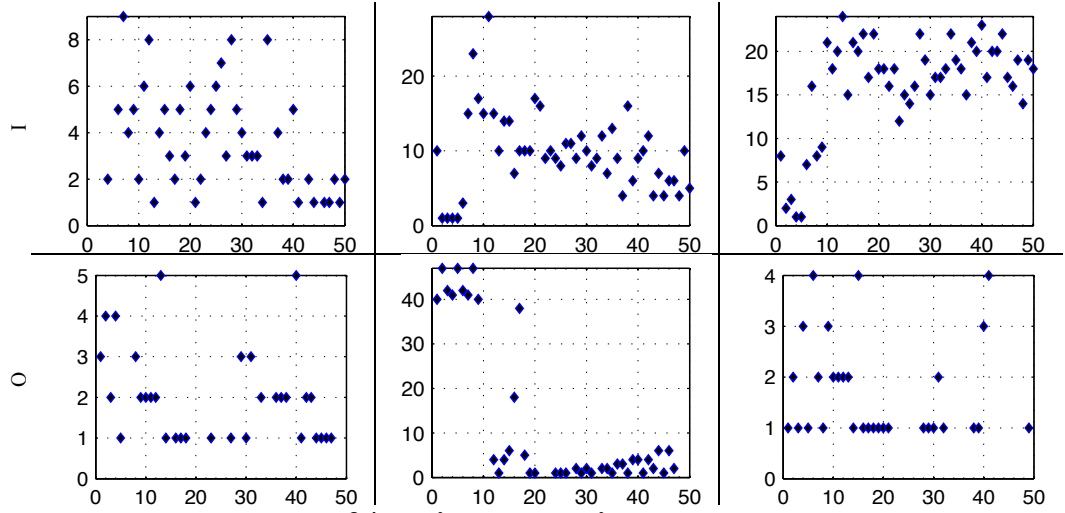

## 24. active_connection_reuse

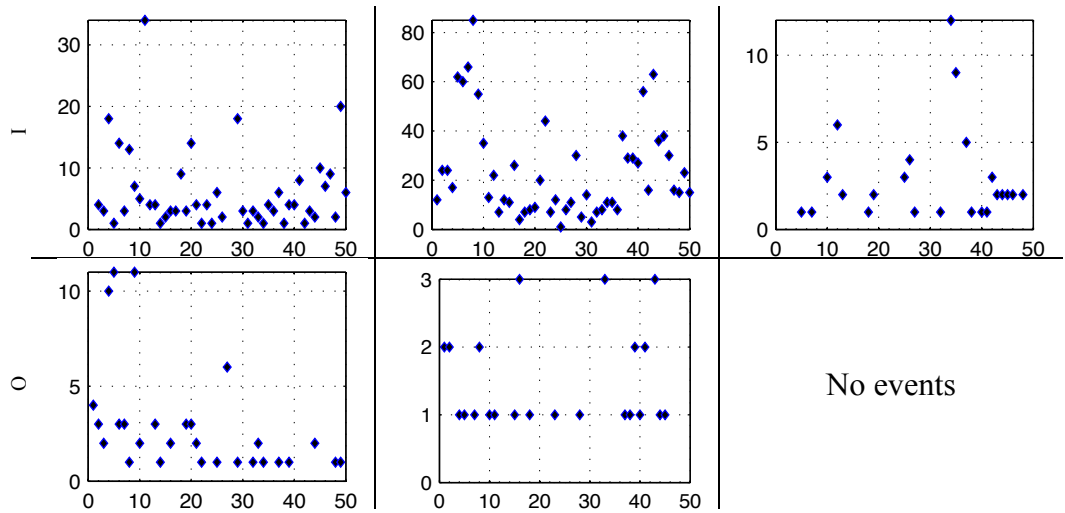

No events

x-axis: 50 periods, y-axis: Number of anomalies found on that period. Left: NET-1, center: NET-2, right: NET-3

## 25. HTTP_chunked_transfer_for_multipart_message

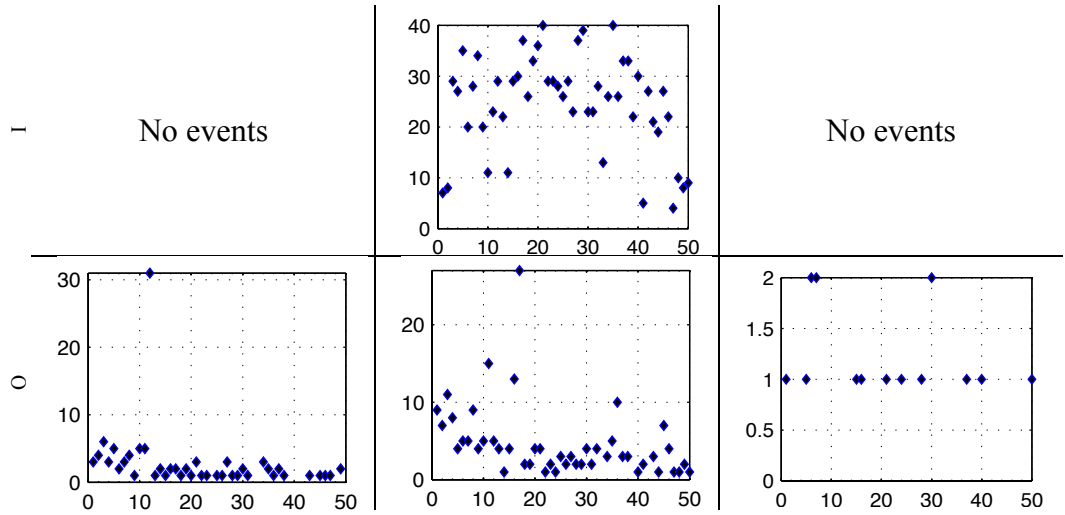

## 26. inappropriate_FIN

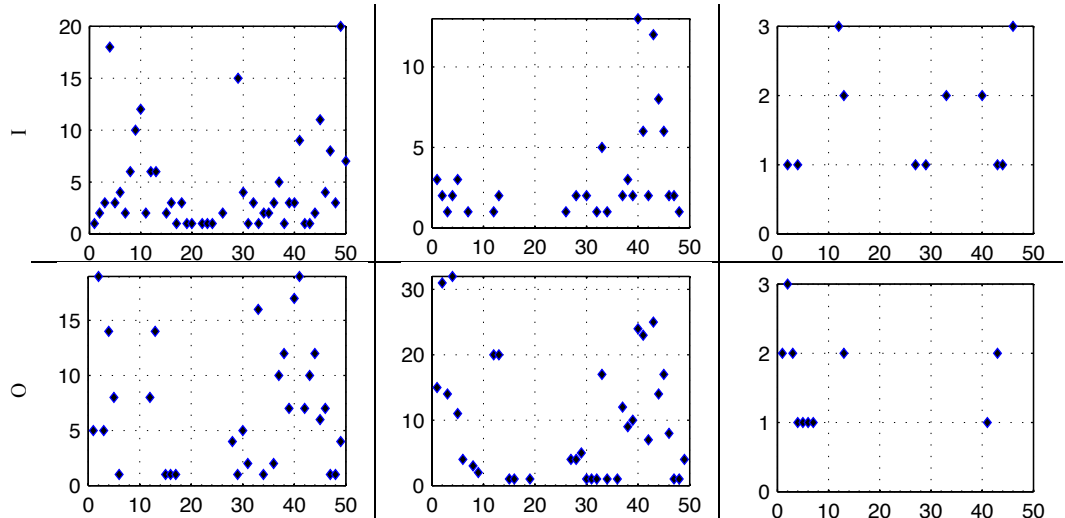

## 27. SYN_after_reset

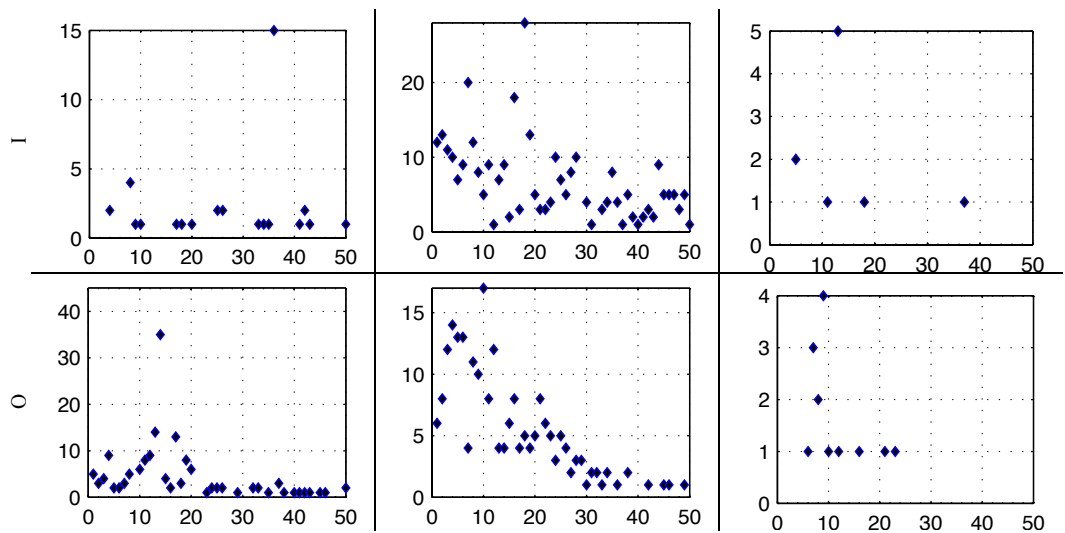

x-axis: 50 periods, y-axis: Number of anomalies found on that period. Left: NET-1, center: NET-2, right: NET-3

## 28. fragment_overlap

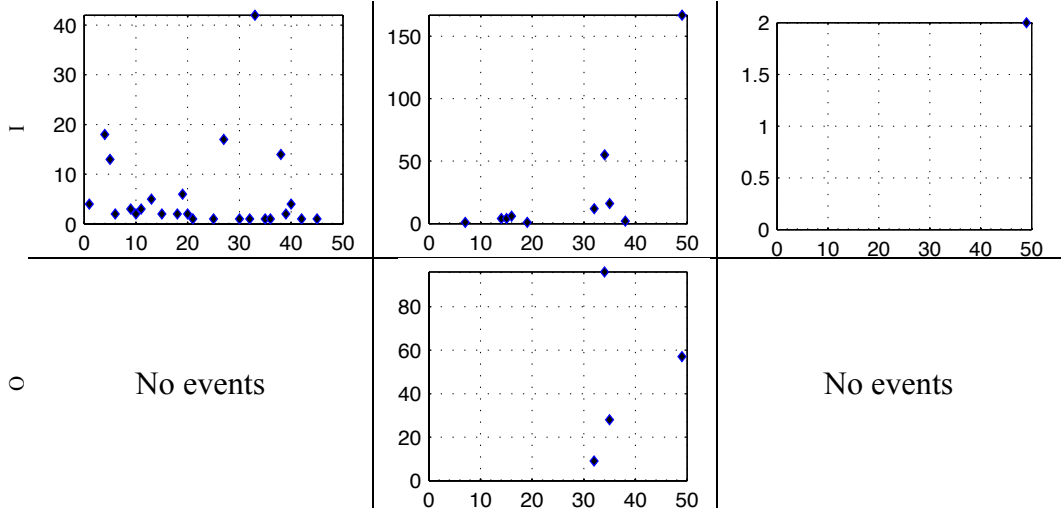

## 29. FIN_advanced_last_seq

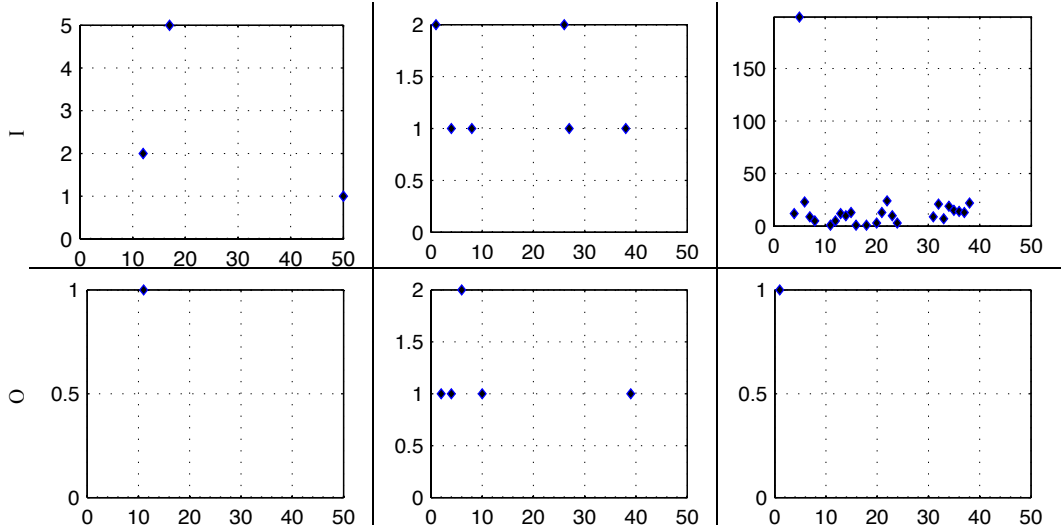

## 30. SYN_inside_connection

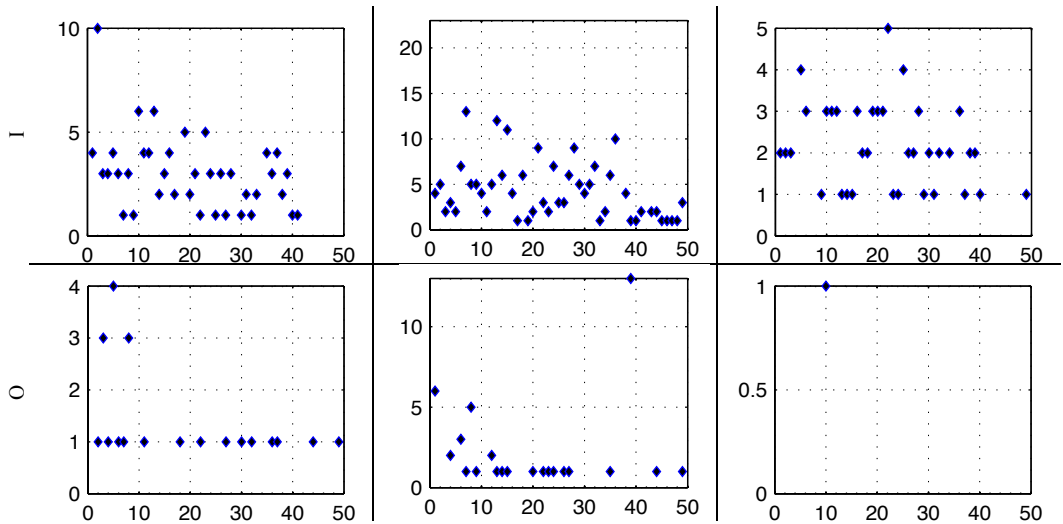

x-axis: 50 periods, y-axis: Number of anomalies found on that period. Left: NET-1, center: NET-2, right: NET-3

### 31. multiple_login_prompts

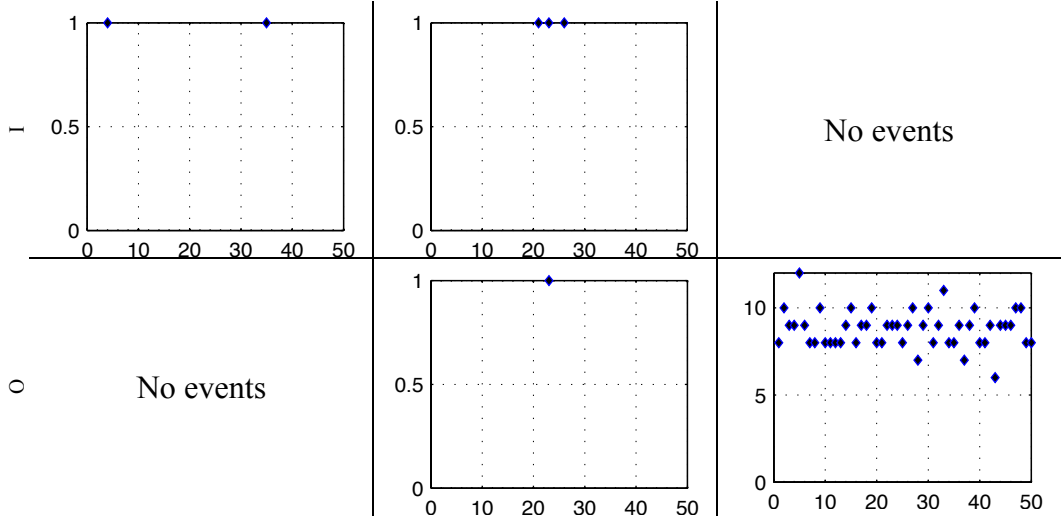

### 32. premature_connection_reuse

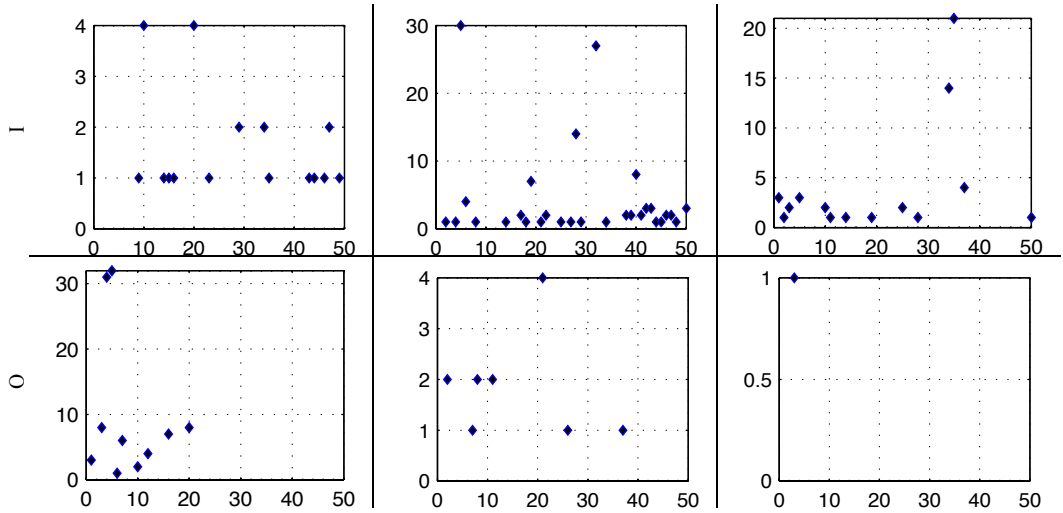

### 33. double_%_in_URI

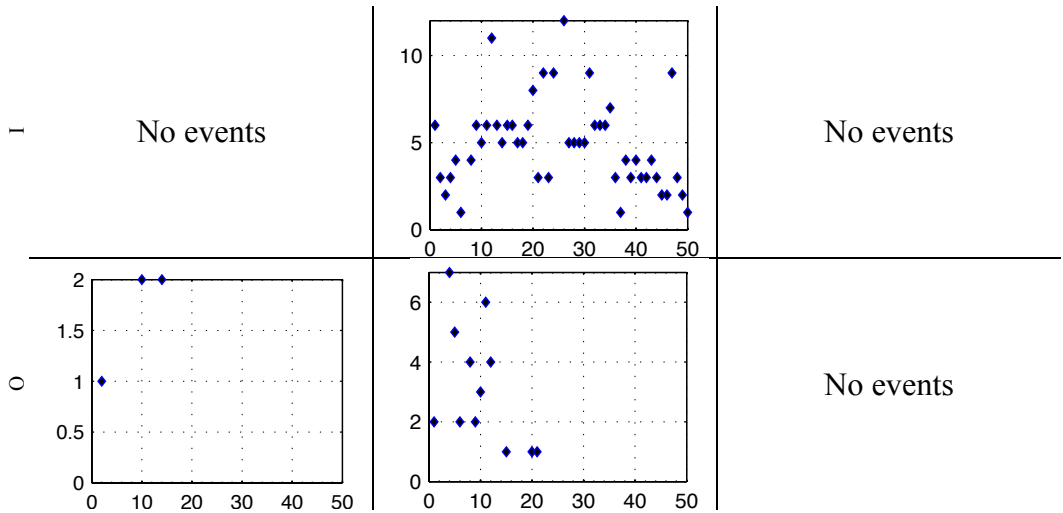

x-axis: 50 periods, y-axis: Number of anomalies found on that period. Left: NET-1, center: NET-2, right: NET-3

## 34. excessive_data_without_further_acks

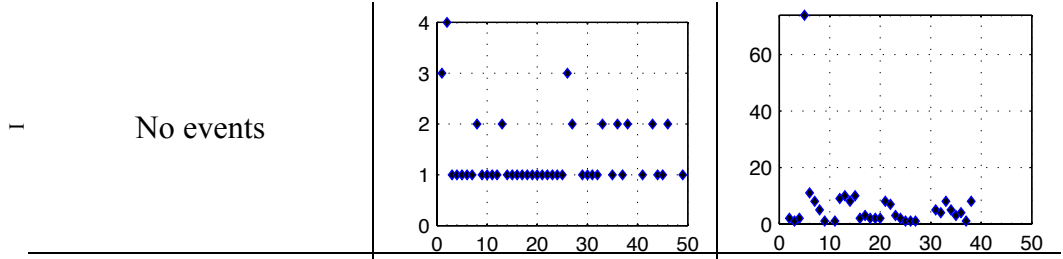

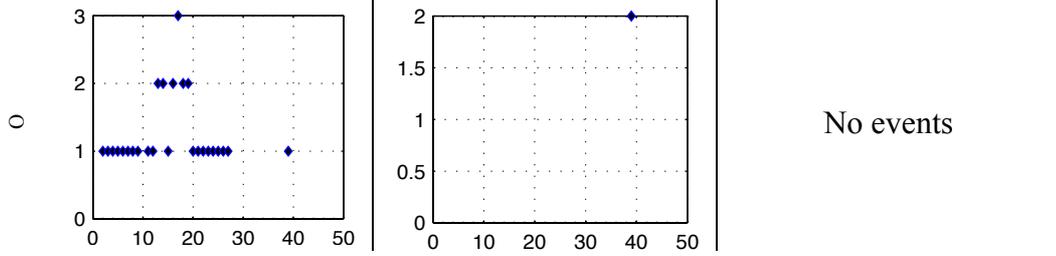

## 35. fragment_with_DF

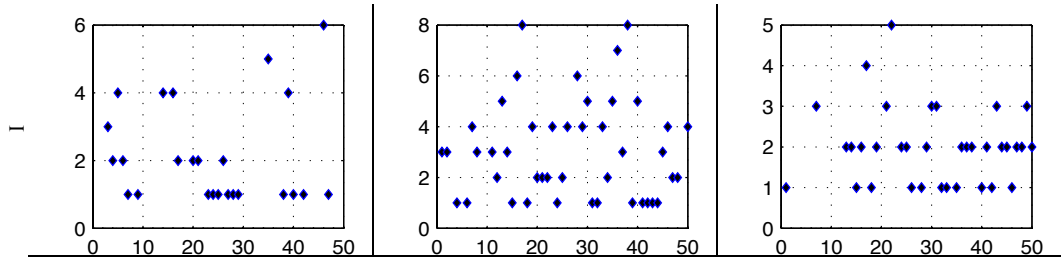

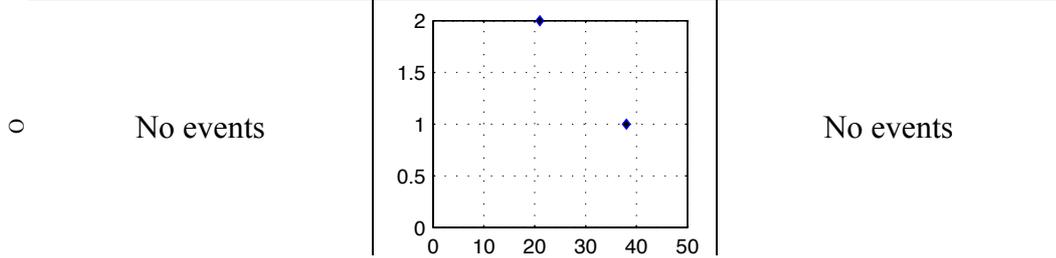

## 36. SYN_after_partial

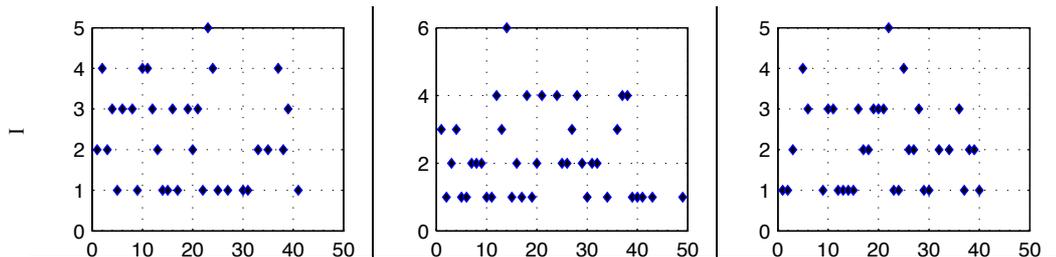

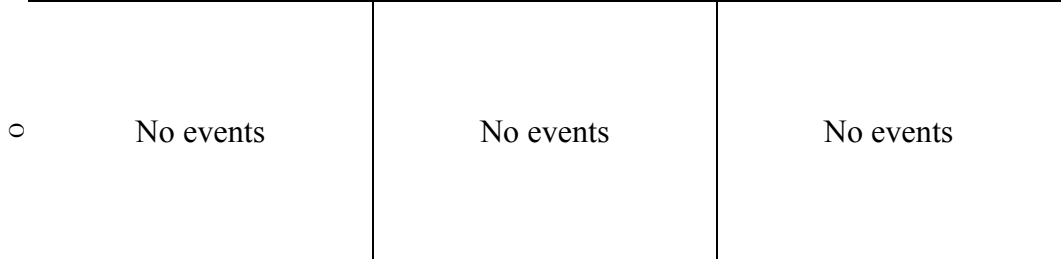

x-axis: 50 periods, y-axis: Number of anomalies found on that period. Left: NET-1, center: NET-2, right: NET-3

## 37. malformed_ssh_identification

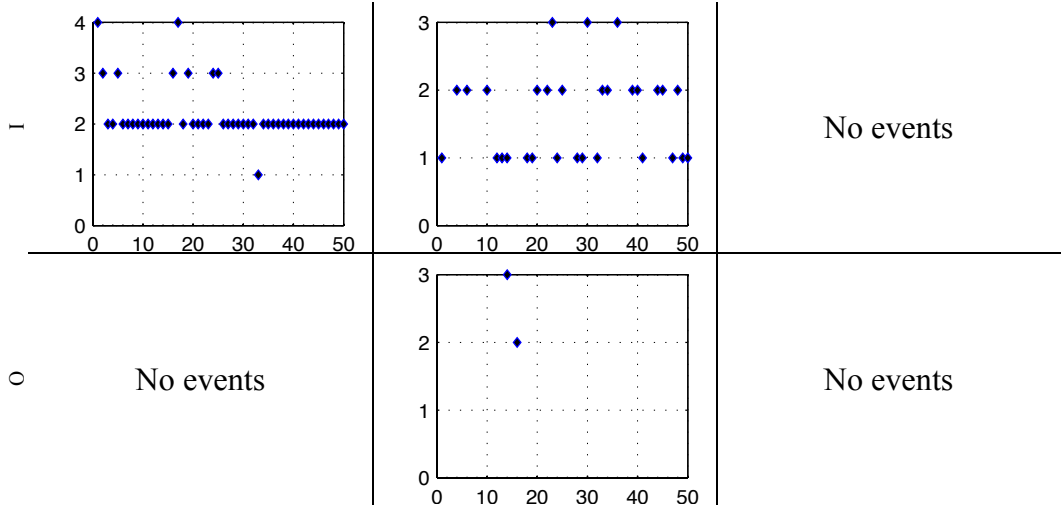

## 38. TCP_christmas

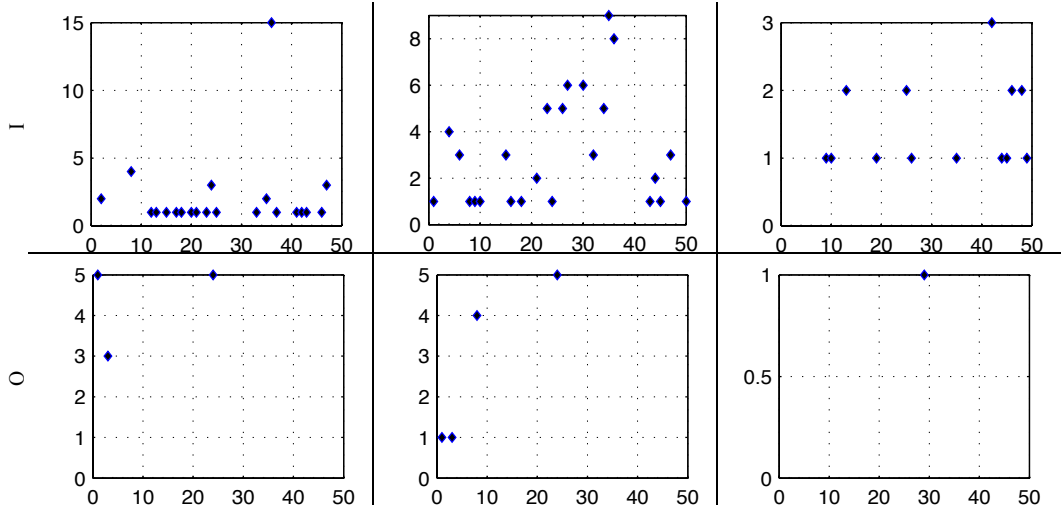

## 39. corrupt_tcp_options

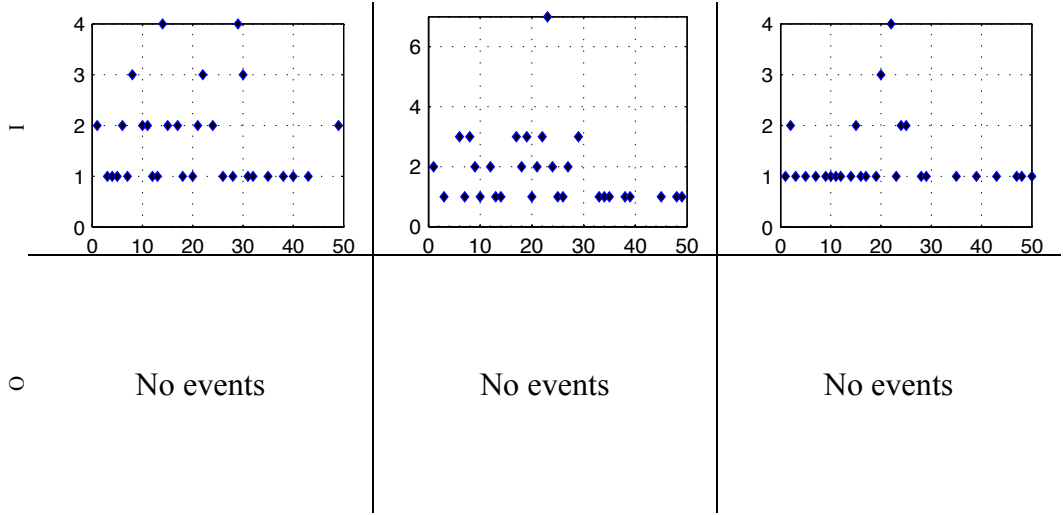

x-axis: 50 periods, y-axis: Number of anomalies found on that period. Left: NET-1, center: NET-2, right: NET-3

## 40. FIN_storm

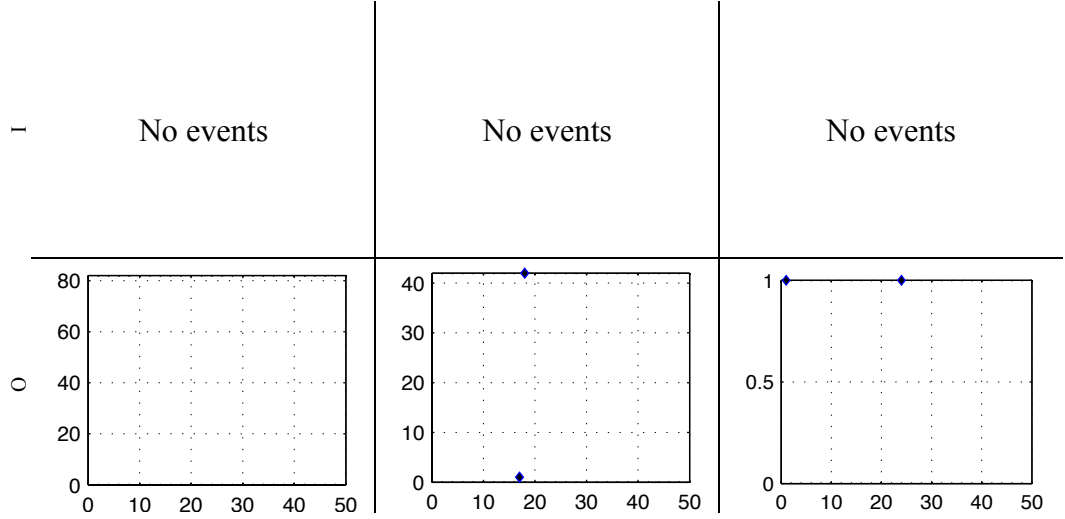

## 41. bad_SYN_ack

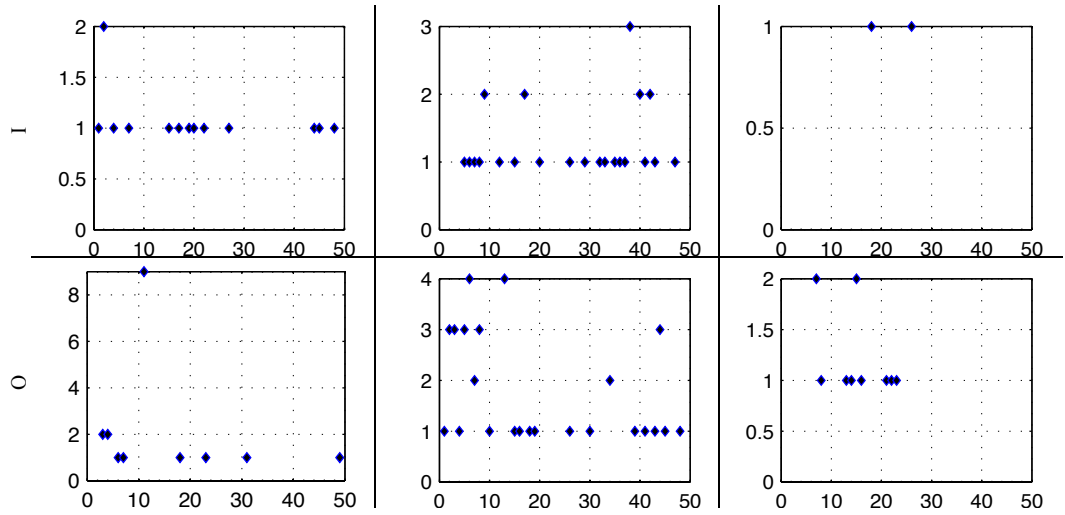

## 42. illegal_%_at_end_of_URI

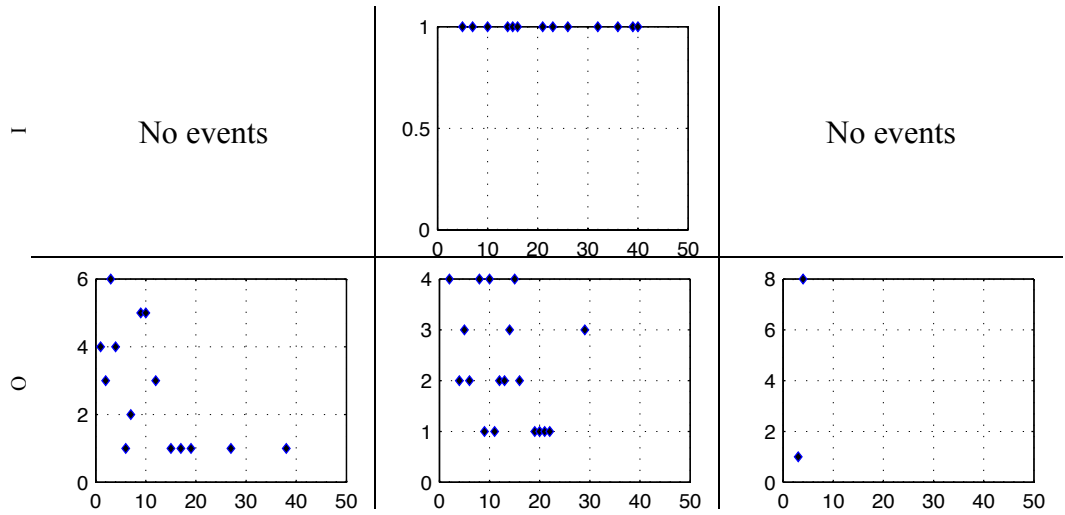

x-axis: 50 periods, y-axis: Number of anomalies found on that period. Left: NET-1, center: NET-2, right: NET-3

### 43. irc_invalid_line

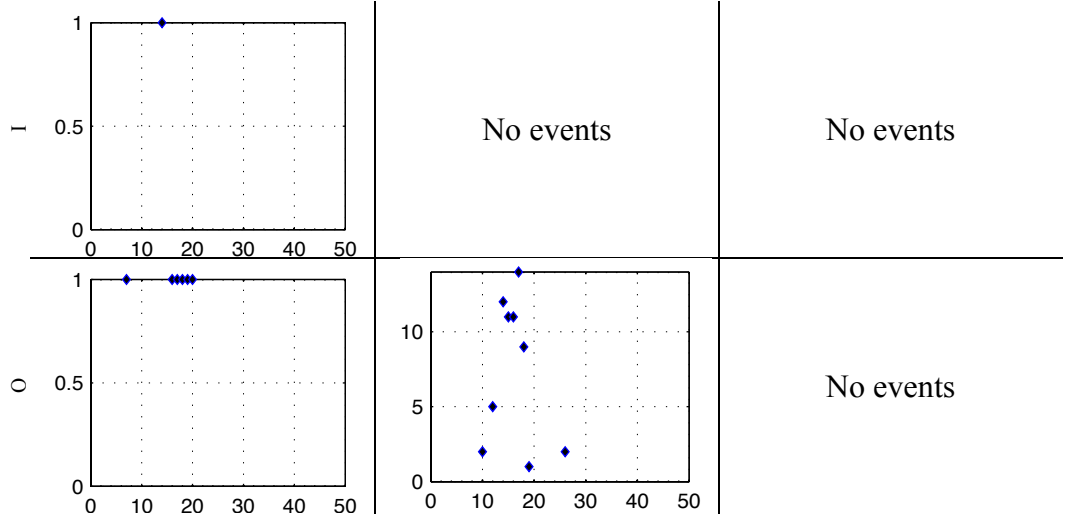

### 44. RST_storm

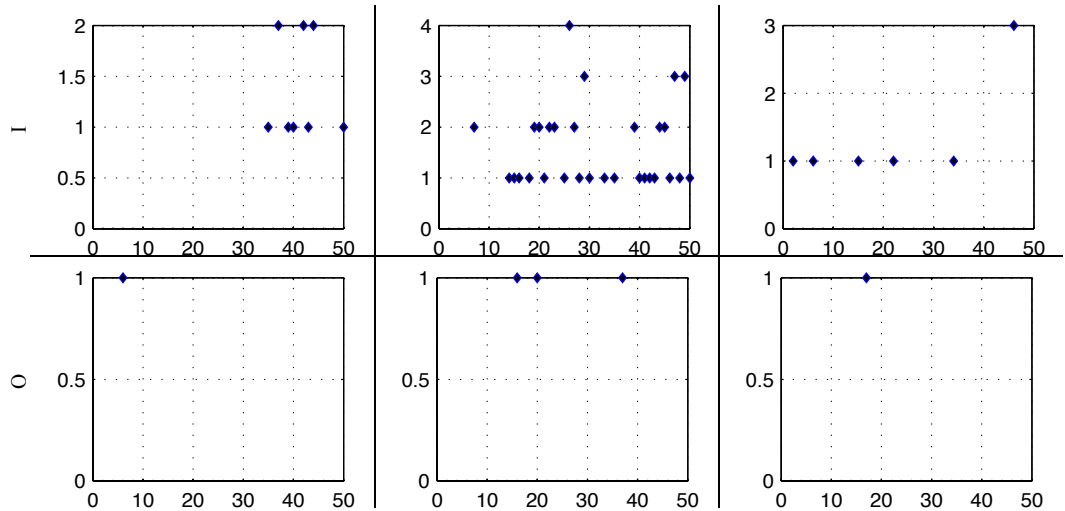

### 45. SYN_after_close

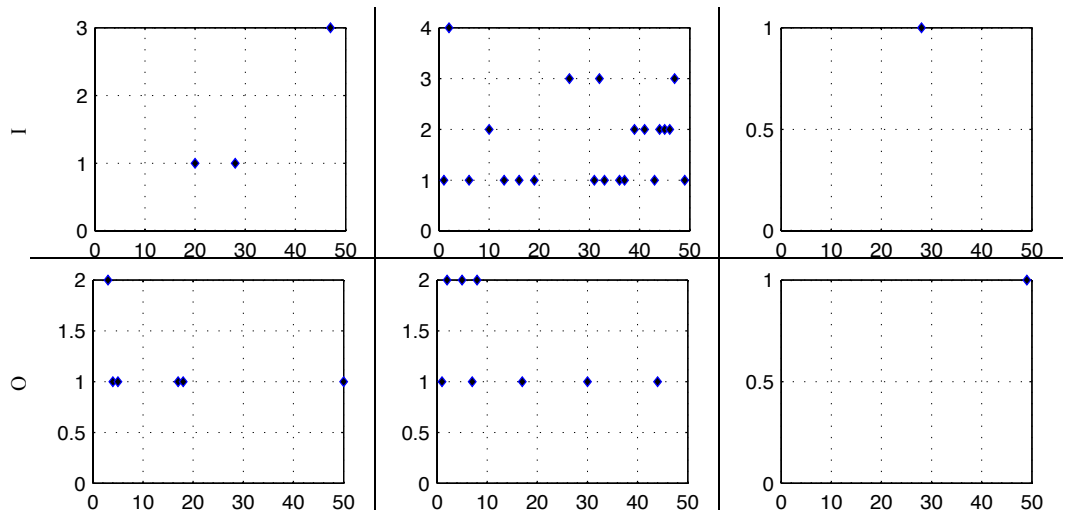

x-axis: 50 periods, y-axis: Number of anomalies found on that period. Left: NET-1, center: NET-2, right: NET-3

## 46. fragment_size_inconsistency

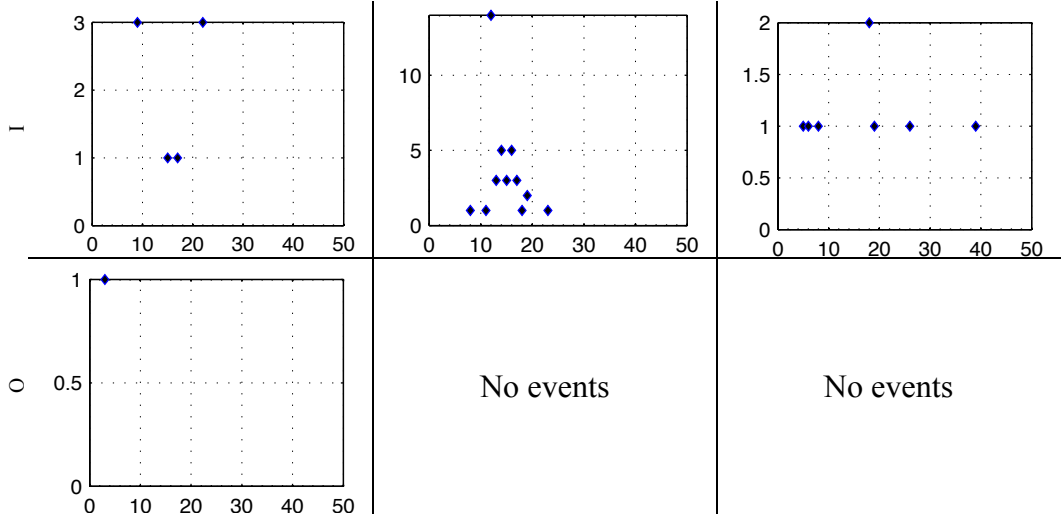

## 47. excessively_small_fragment

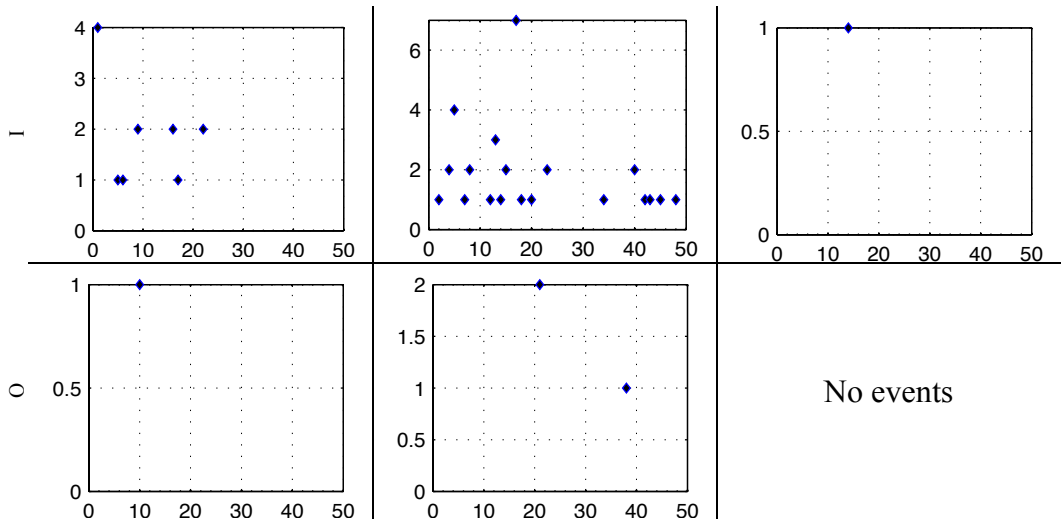

## 48. simultaneous_open

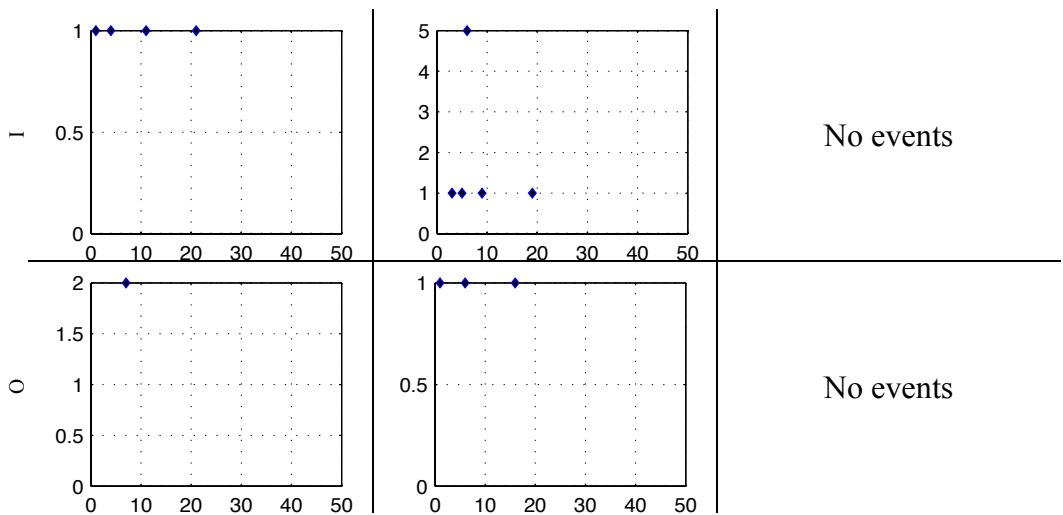

x-axis: 50 periods, y-axis: Number of anomalies found on that period. Left: NET-1, center: NET-2, right: NET-3

### 49. bad_TCP_header_len

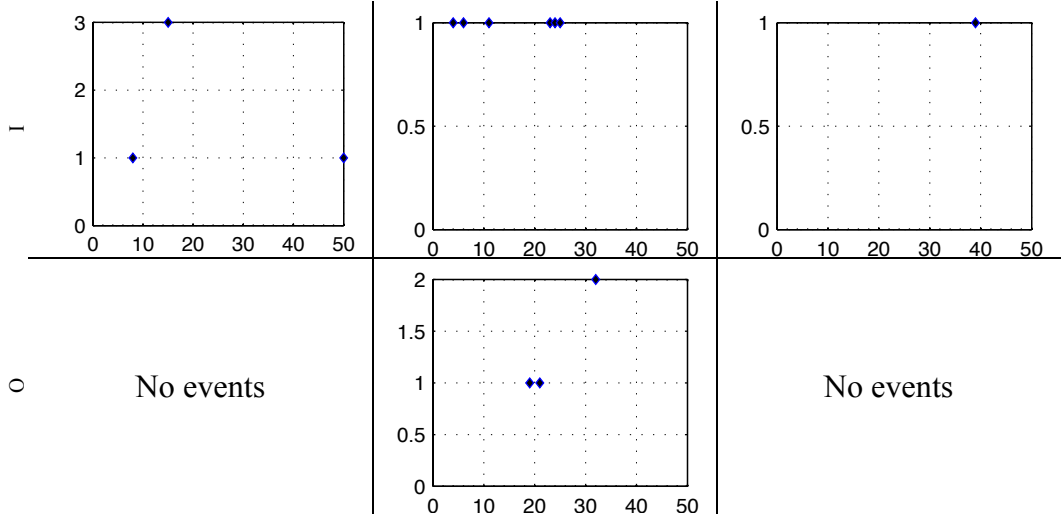

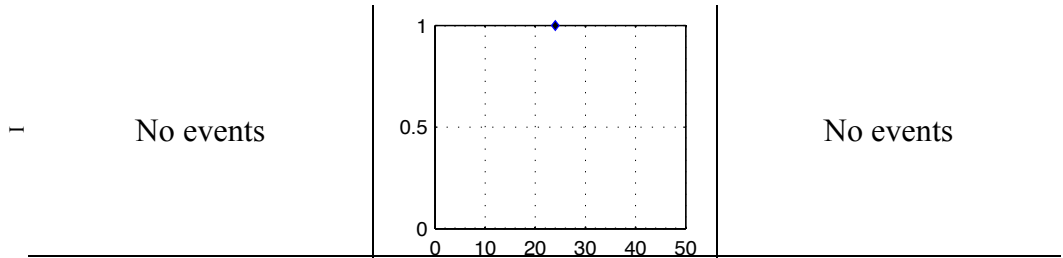

### 50. irc_invalid_command

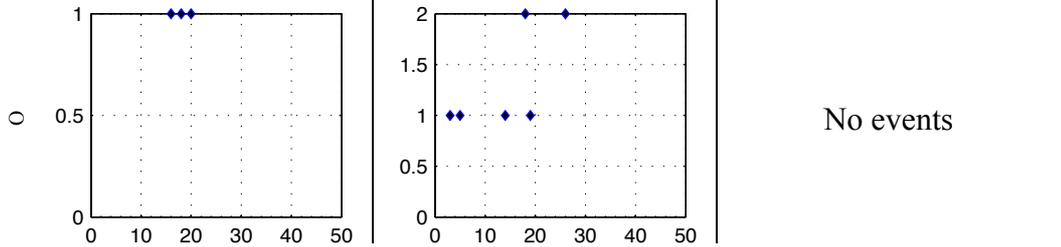

### 51. excessively_large_fragment

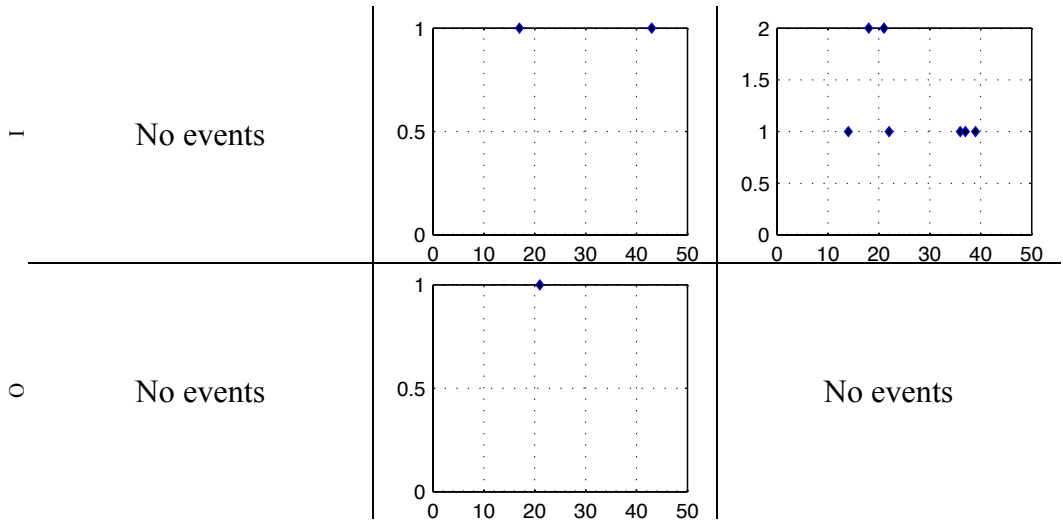

x-axis: 50 periods, y-axis: Number of anomalies found on that period. Left: NET-1, center: NET-2, right: NET-3

## 52. HTTP_bad_chunk_size

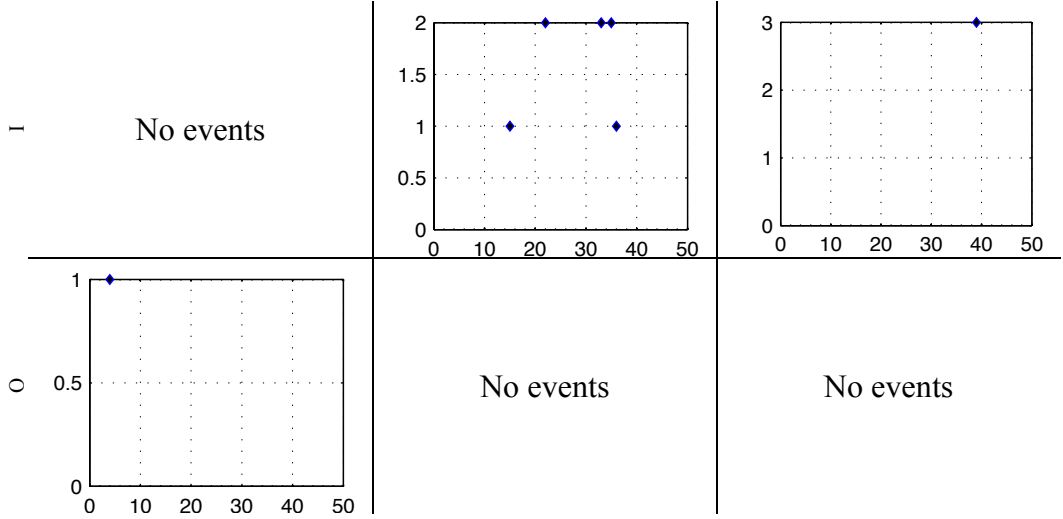

## 53. unpaired_RPC_response

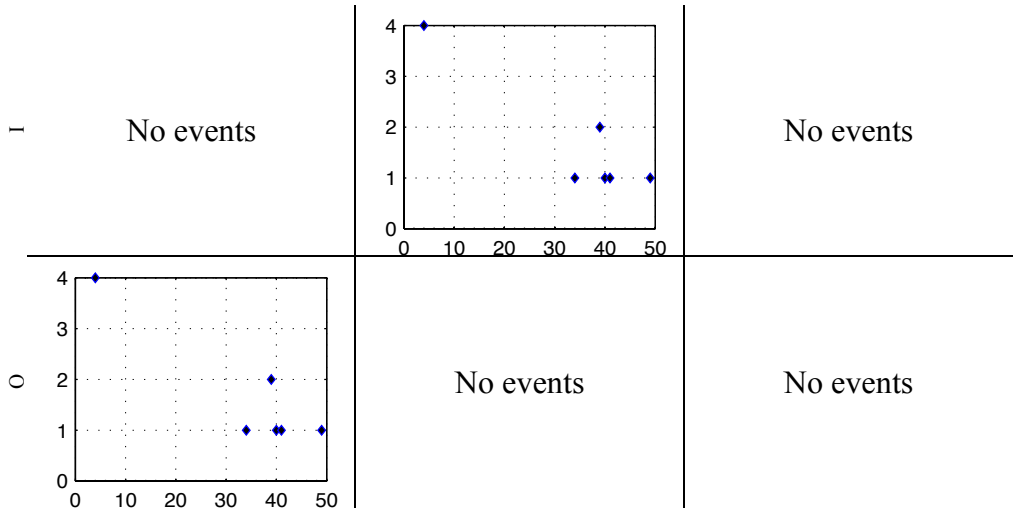

## 54. irc_invalid_topic_reply

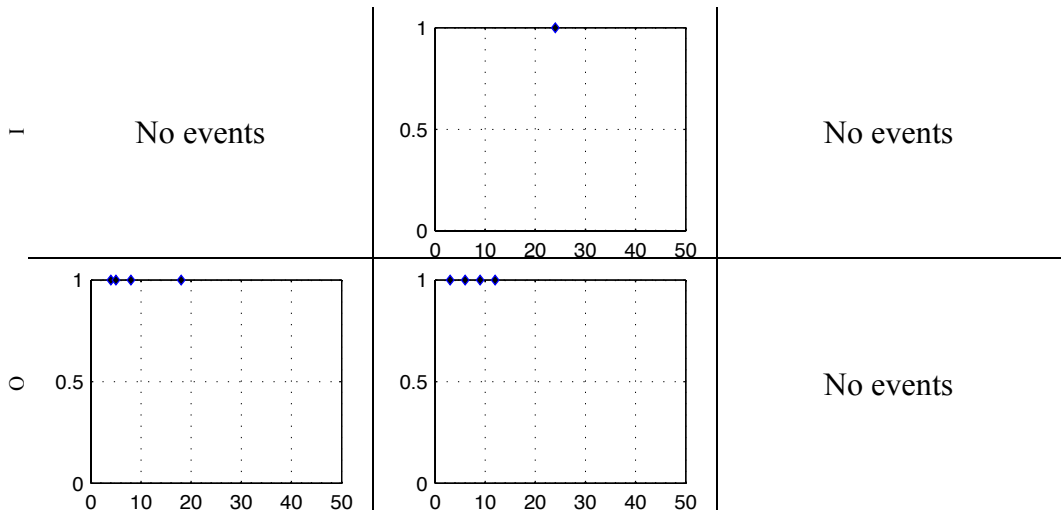

x-axis: 50 periods, y-axis: Number of anomalies found on that period. Left: NET-1, center: NET-2, right: NET-3

**55. baroque_SYN**

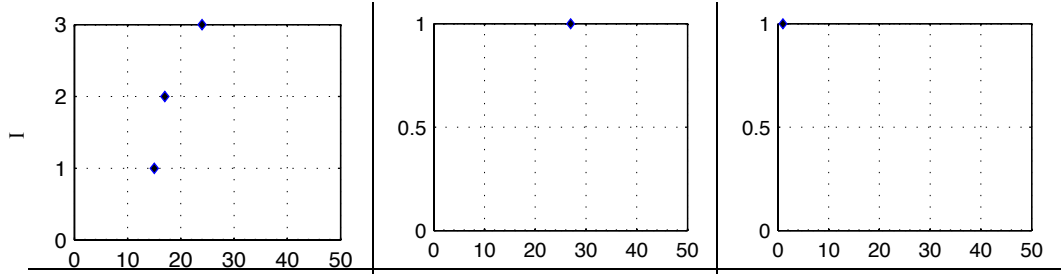

O        No events                    No events                    No events

**56. no_login_prompt**

–        No events                    No events                    No events

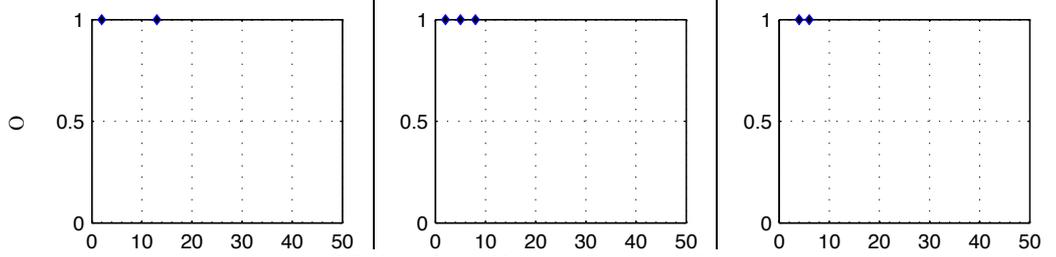

**57. irc_invalid_reply_number**

–        No events                    No events                    No events

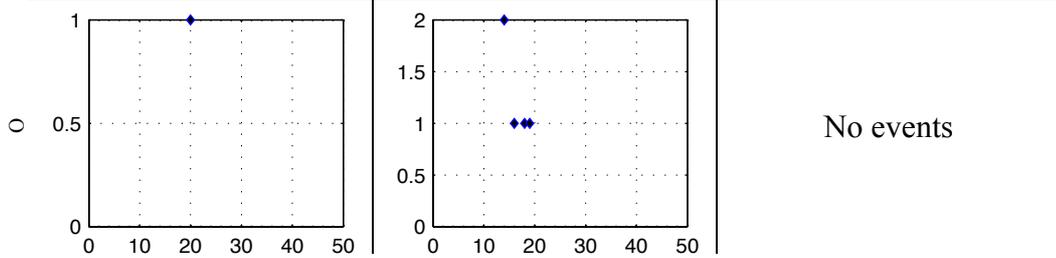

x-axis: 50 periods, y-axis: Number of anomalies found on that period. Left: NET-1, center: NET-2, right: NET-3

## 58. partial_RPC_request

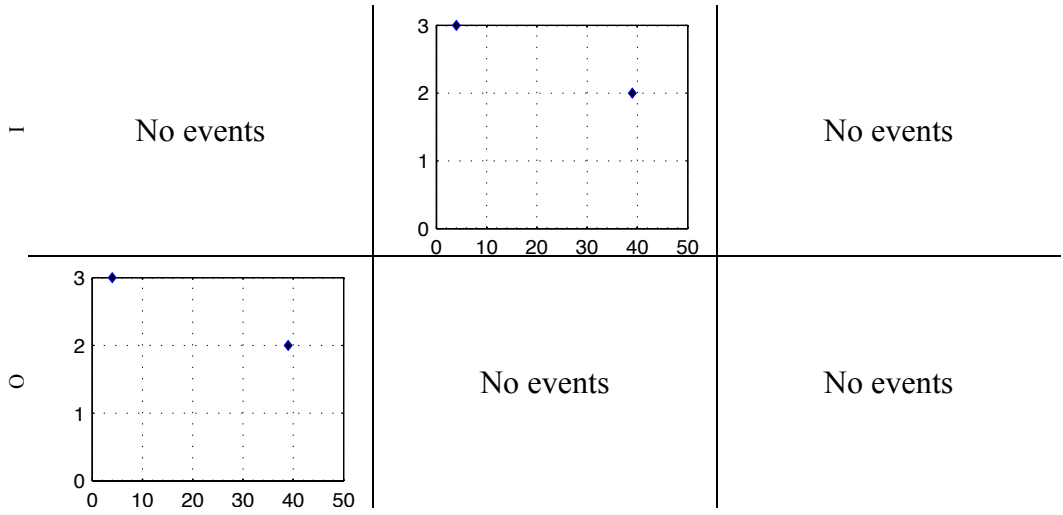

## 59. repeated_SYN_with_ACK

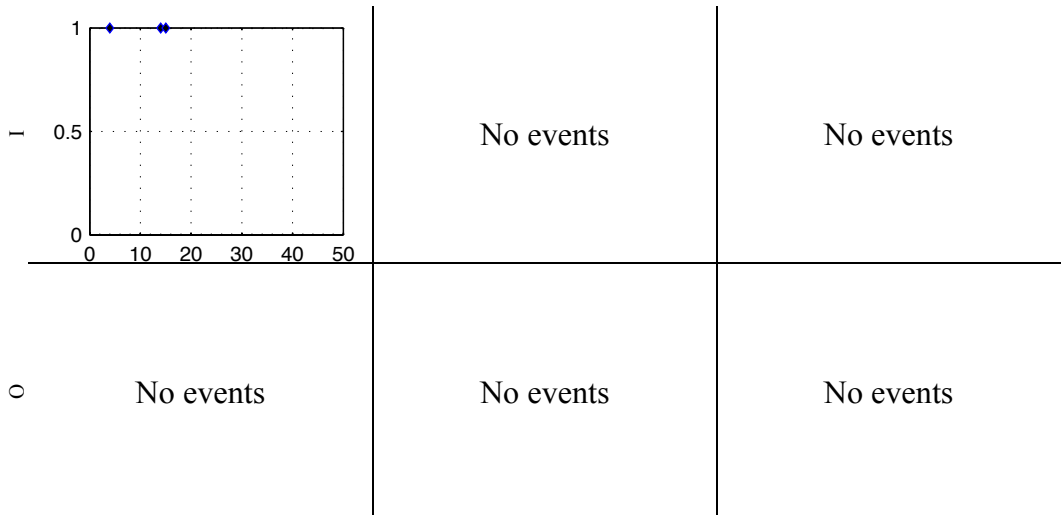

## 60. partial_ftp_request

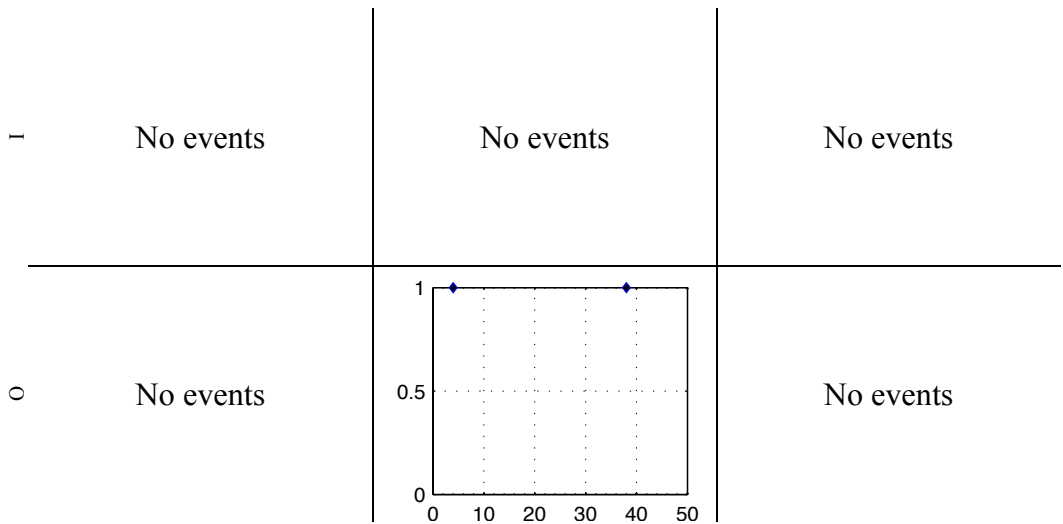

x-axis: 50 periods, y-axis: Number of anomalies found on that period. Left: NET-1, center: NET-2, right: NET-3

## 61. irc_line_size_exceeded

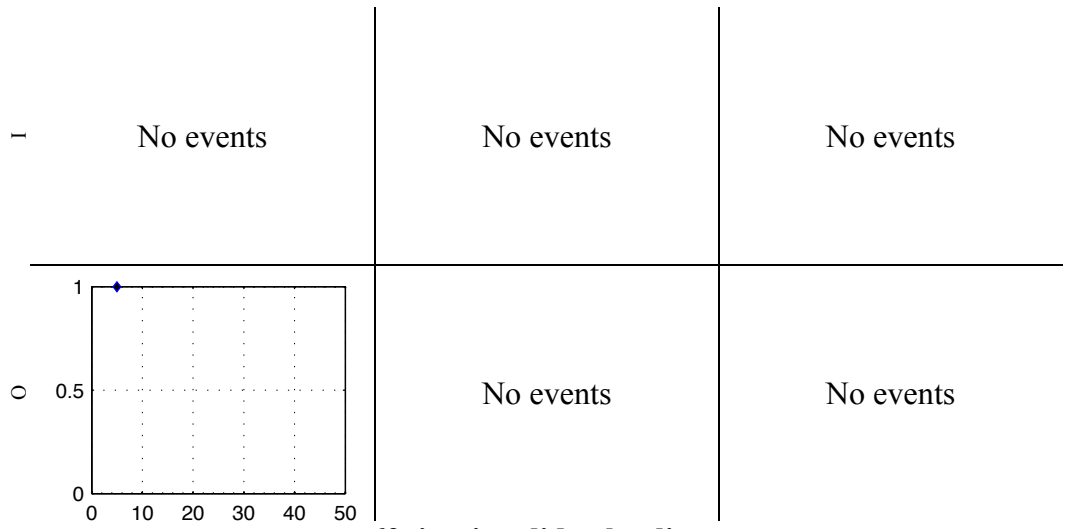

## 62. irc_invalid_who_line

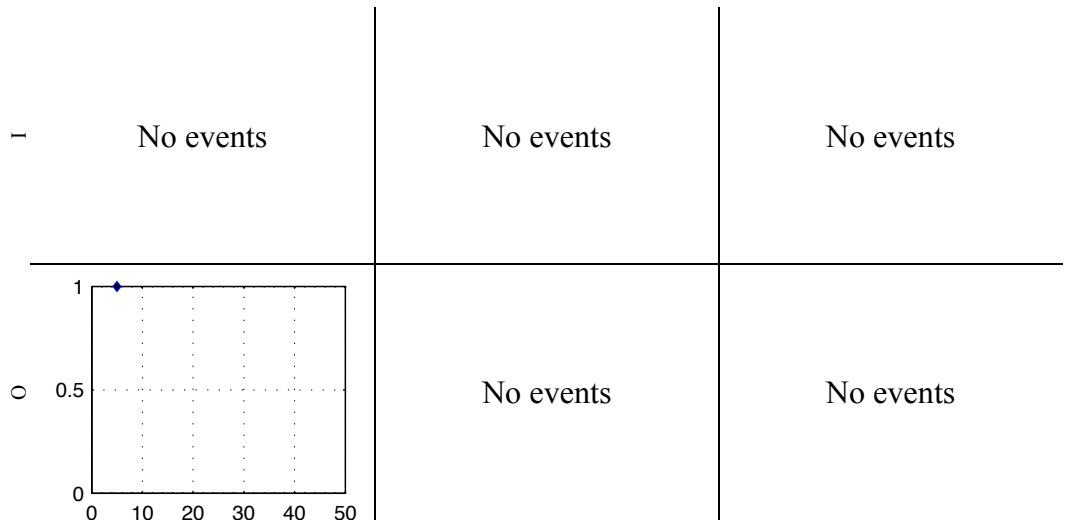

x-axis: 50 periods, y-axis: Number of anomalies found on that period. Left: NET-1, center: NET-2, right: NET-3

**Appendix**

**B   Reference for the analyzed periods**

**Analyzed Periods**

*15$^{th}$ June 2009*

| Id | From | To |
|----|------|-----|
| 1 | 19:24:49 | 19:34:48 |
| 2 | 19:35:47 | 19:45:46 |
| 3 | 19:46:44 | 19:56:43 |
| 4 | 19:57:48 | 20:07:47 |
| 5 | 20:09:00 | 20:19:59 |
| 6 | 20:20:43 | 20:30:42 |
| 7 | 20:32:39 | 20:42:38 |
| 8 | 20:45:00 | 20:55:59 |
| 9 | 20:56:41 | 21:06:40 |
| 10 | 21:09:51 | 21:19:50 |
| 11 | 21:23:40 | 21:33:39 |
| 12 | 21:36:57 | 21:46:56 |
| 13 | 21:50:51 | 22:00:50 |
| 14 | 22:03:07 | 22:13:06 |
| 15 | 22:15:56 | 22:25:55 |
| 16 | 22:29:33 | 22:39:32 |
| 17 | 22:42:33 | 22:52:32 |
| 18 | 22:54:56 | 23:04:55 |
| 19 | 23:07:19 | 23:17:18 |
| 20 | 23:20:21 | 23:30:20 |
| 21 | 23:34:41 | 23:44:40 |
| 22 | 23:50:51 | 0:00:50 |

*16$^{th}$ June 2009*

| Id | From | To |
|----|------|-----|
| 23 | 0:09:43 | 0:19:42 |
| 24 | 0:19:43 | 0:29:42 |
| 25 | 0:30:08 | 0:40:07 |
| 26 | 0:40:08 | 0:50:07 |
| 27 | 0:50:25 | 1:00:24 |
| 28 | 1:06:46 | 1:16:45 |
| 29 | 1:21:07 | 1:31:06 |
| 30 | 1:37:27 | 1:47:26 |
| 31 | 1:52:20 | 2:02:19 |
| 32 | 2:07:04 | 2:17:03 |
| 33 | 2:32:01 | 2:42:00 |
| 34 | 2:42:01 | 2:52:00 |
| 35 | 2:52:01 | 3:02:00 |
| 36 | 3:04:58 | 3:14:57 |
| 37 | 3:20:12 | 3:30:11 |
| 38 | 3:31:00 | 3:40:59 |
| 39 | 4:01:50 | 4:11:49 |
| 40 | 4:16:11 | 4:26:10 |
| 41 | 4:26:43 | 4:36:42 |
| 42 | 4:37:35 | 4:47:34 |
| 43 | 4:48:52 | 4:58:51 |
| 44 | 5:00:18 | 5:10:17 |

*16<sup>th</sup> June 2009*

| Id | From | To |
|----|------|-----|
| 45 | 5:11:40 | 5:21:39 |
| 46 | 5:23:31 | 5:33:30 |
| 47 | 5:40:12 | 5:50:11 |
| 48 | 5:50:12 | 6:00:11 |
| 49 | 6:00:12 | 6:10:11 |
| 50 | 6:13:01 | 6:23:00 |